\documentclass[twoside,twocolumn,9pt]{article}
\usepackage{extsizes}
\usepackage[super,sort&compress,comma]{natbib} 
\usepackage[version=3]{mhchem}
\usepackage[left=1.5cm, right=1.5cm, top=1.785cm, bottom=2.0cm]{geometry}
\usepackage{balance}
\usepackage{mathptmx}
\usepackage{sectsty}
\usepackage{graphicx} 
\usepackage{lastpage}
\usepackage[format=plain,justification=justified,singlelinecheck=false,font={stretch=1.125,small,sf},labelfont=bf,labelsep=space]{caption}
\usepackage{float}
\usepackage{fancyhdr}
\usepackage{fnpos}
\usepackage[english]{babel}
\addto{\captionsenglish}{%
  
}
\usepackage{array}
\usepackage{droidsans}
\usepackage{charter}
\usepackage[T1]{fontenc}
\usepackage[usenames,dvipsnames]{xcolor}
\usepackage{setspace}
\usepackage[compact]{titlesec}
\usepackage{hyperref}

\usepackage{epstopdf}

\definecolor{cream}{RGB}{222,217,201}

\begin{document}

\pagestyle{fancy}
\thispagestyle{plain}
\fancypagestyle{plain}{
\renewcommand{\headrulewidth}{0pt}
}

\makeFNbottom
\makeatletter
\renewcommand\LARGE{\@setfontsize\LARGE{15pt}{17}}
\renewcommand\Large{\@setfontsize\Large{12pt}{14}}
\renewcommand\large{\@setfontsize\large{10pt}{12}}
\renewcommand\footnotesize{\@setfontsize\footnotesize{7pt}{10}}
\makeatother

\renewcommand{\thefootnote}{\fnsymbol{footnote}}
\renewcommand\footnoterule{\vspace*{1pt}%
\color{cream}\hrule width 3.5in height 0.4pt \color{black}\vspace*{5pt}} 
\setcounter{secnumdepth}{5}

\makeatletter 
\renewcommand\@biblabel[1]{#1}            
\renewcommand\@makefntext[1]%
{\noindent\makebox[0pt][r]{\@thefnmark\,}#1}
\makeatother 
\renewcommand{\figurename}{\small{Fig.}~}
\sectionfont{\sffamily\Large}
\subsectionfont{\normalsize}
\subsubsectionfont{\bf}
\setstretch{1.125} 
\setlength{\skip\footins}{0.8cm}
\setlength{\footnotesep}{0.25cm}
\setlength{\jot}{10pt}
\titlespacing*{\section}{0pt}{4pt}{4pt}
\titlespacing*{\subsection}{0pt}{15pt}{1pt}

\fancyfoot{}
\fancyfoot[LO,RE]{\vspace{-7.1pt}\includegraphics[height=9pt]{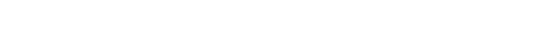}}
\fancyfoot[CO]{\vspace{-7.1pt}\hspace{13.2cm}\includegraphics{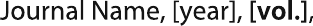}}
\fancyfoot[CE]{\vspace{-7.2pt}\hspace{-14.2cm}\includegraphics{head_foot/RF}}
\fancyfoot[RO]{\footnotesize{\sffamily{1--\pageref{LastPage} ~\textbar  \hspace{2pt}\thepage}}}
\fancyfoot[LE]{\footnotesize{\sffamily{\thepage~\textbar\hspace{3.45cm} 1--\pageref{LastPage}}}}
\fancyhead{}
\renewcommand{\headrulewidth}{0pt} 
\renewcommand{\footrulewidth}{0pt}
\setlength{\arrayrulewidth}{1pt}
\setlength{\columnsep}{6.5mm}
\setlength\bibsep{1pt}

\makeatletter 
\newlength{\figrulesep} 
\setlength{\figrulesep}{0.5\textfloatsep} 

\newcommand{\topfigrule}{\vspace*{-1pt}%
\noindent{\color{cream}\rule[-\figrulesep]{\columnwidth}{1.5pt}} }

\newcommand{\botfigrule}{\vspace*{-2pt}%
\noindent{\color{cream}\rule[\figrulesep]{\columnwidth}{1.5pt}} }

\newcommand{\dblfigrule}{\vspace*{-1pt}%
\noindent{\color{cream}\rule[-\figrulesep]{\textwidth}{1.5pt}} }

\makeatother

\twocolumn[
  \begin{@twocolumnfalse}
{\includegraphics[height=30pt]{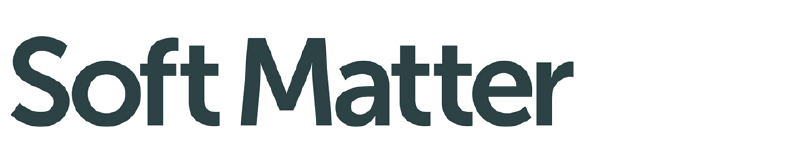}\hfill\raisebox{0pt}[0pt][0pt]{\includegraphics[height=55pt]{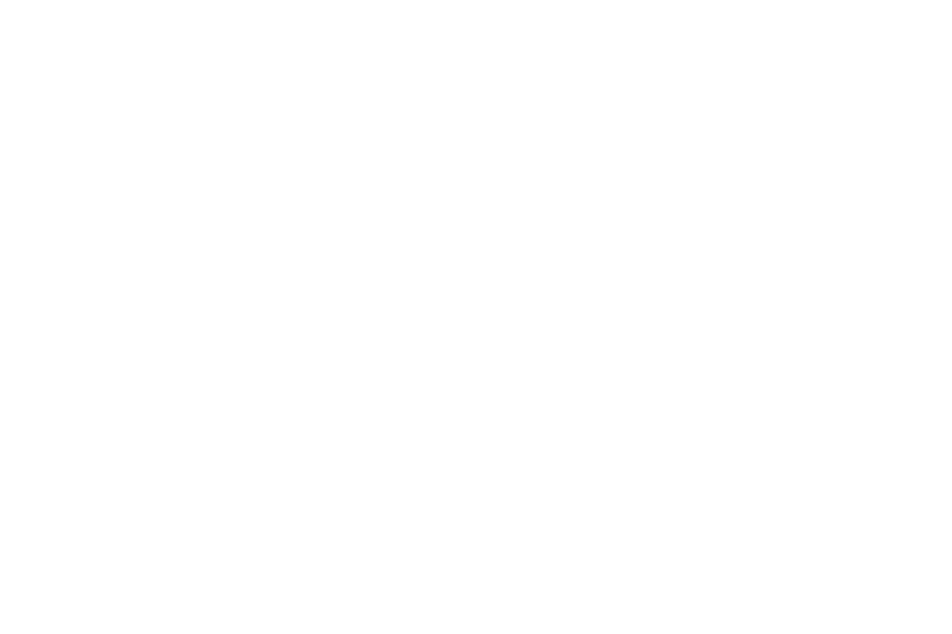}}\\[1ex]
\includegraphics[width=18.5cm]{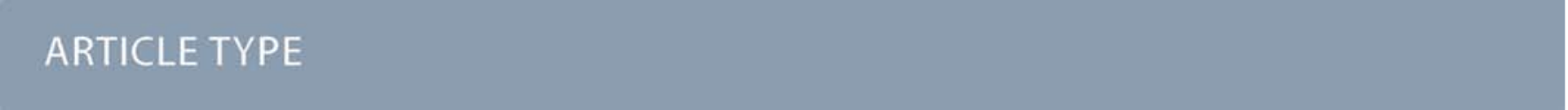}}\par
\vspace{1em}
\sffamily
\begin{tabular}{m{4.5cm} p{13.5cm} }

\includegraphics{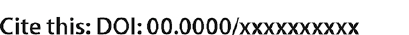} & \noindent\LARGE{\textbf{Role of Entropy in Determining the Phase Behavior of Protein Solutions Induced by Multivalent Ions$^\dag$}} \\
\vspace{0.3cm} & \vspace{0.3cm} \\

	& \noindent\large{Anil Kumar Sahoo,$^{\ast}$\textit{$^{abc}$} Frank Schreiber,\textit{$^{d}$} Roland R. Netz,\textit{$^{ce}$} and Prabal K. Maiti$^{\ast}$\textit{$^{a}$}} \\

\includegraphics{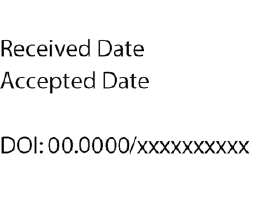} & \noindent\normalsize{
Recent experiments have reported lower critical solution temperature (LCST) phase behavior of aqueous solutions of proteins
induced by multivalent ions, where the solution phase separates upon heating. This phenomenon is linked to complex hydration
effects that result in a net entropy gain upon phase separation. 
	To decipher the underlying molecular mechanism, we use all-atom molecular dynamics simulations along with the two-phase
thermodynamic method for entropy calculation. 
	Based on simulations of a single BSA protein in various salt solutions (NaCl, CaCl$_2$, MgCl$_2$,
and YCl$_3$) at temperatures ($T$) ranging 283--323 K, we find that the cation--protein binding affinity \textit{increases}
with $T$, reflecting its thermodynamic driving force to be entropic in origin.
We show that in the cation binding process, many tightly bound water molecules from the solvation shells of a cation and
the protein are released to the bulk, resulting in entropy gain.
	To rationalize the LCST behavior, we calculate the $\zeta$-potential that shows \textit{charge inversion} of the protein
for solutions containing multivalent ions. The $\zeta$-potential increases with $T$.
Performing simulations of two BSA proteins, we demonstrate that the protein--protein binding is mediated by multiple
cation bridges and involves similar dehydration effects that cause a large entropy gain which more than compensates for rotational
and translational entropy losses of the proteins.
	Thus, the LCST behavior is entropy-driven, but the associated solvation effects are markedly different from
hydrophobic hydration. Our findings have direct implications for tuning the phase behavior of
biological and soft-matter systems, \textit{e.g.}, protein condensation and crystallization.
} \\

\end{tabular}

 \end{@twocolumnfalse} \vspace{0.6cm}

  ]

\renewcommand*\rmdefault{bch}\normalfont\upshape
\rmfamily
\section*{}
\vspace{-1cm}


\footnotetext{\textit{$^{a}$~Center for Condensed Matter Theory, Department of Physics, Indian Institute of Science, Bangalore-560012, India. E-mail: 201992kumarsahoo@gmail.com, maiti@iisc.ac.in}}
\footnotetext{\textit{$^{b}$~Max Planck Institute of Colloids and Interfaces, Am M\"uhlenberg 1, 14476 Potsdam, Germany.}}
\footnotetext{\textit{$^{c}$~Fachbereich Physik, Freie Universit\"at Berlin, Arnimallee 14, 14195 Berlin, Germany.}}
\footnotetext{\textit{$^{d}$~Institute for Applied Physics, University of T\"ubingen, 72076 T\"ubingen, Germany.}}
\footnotetext{\textit{$^{e}$~Department of Physics, Indian Institute of Science, Bangalore-560012, India.}}

\footnotetext{\dag~Electronic Supplementary Information (ESI) available: [Distribution of anions around the protein;
electrostatic potential map and cation/anion number density map; cation binding/unbinding kinetics; 
time series of the number of protein-bound water molecules; radial distribution functions for oxygen of water around cations;
entropy contribution of water in the second solvation shell of cations; comparison of entropy obtained from the 2PT method
and the temperature derivative of free energy; the $T$-dependence of binding energy; $T$ and salt concentration dependence of
the dielectic constant; the $T$-dependence of entropy and energy of ion binding excluding the contribution due to dehydration of
the second solvation shell of the ion; note on the $T$-dependence of free energy; theory of the 2PT method; note on volume calculation;
derivation of the expression for $\zeta$-potential (Eq. \ref{eq:ZetaPot}); table for comparison between computed and experimental values for
structural parameters and entropy of ion hydration; and table for water released from protein and cation surfaces in a binding process]. See DOI: 10.1039/cXsm00000x/}






\section*{Introduction}
Ions play an important role in many biophysical processes, \textit{e.g.}, allosteric regulation, enzymatic activity, DNA condensation,
and protein solubility and crystallization. Starting from the pioneering works by Hofmeister, there has been immense progress made to
better understand ion--protein interactions.\cite{jungwirth2014beyond, okur2017beyond}
In recent years, due to various applications in biology, medicine and physics, there is increasing interest
to tune and control the phase behavior of protein solutions using multivalent ions.\cite{matsarskaia2020multivalent}
Diverse phenomena induced by multivalent ions have been realized in experiments. These include: (i) reentrant condensation of
proteins in bulk solution\cite{zhang2008reentrant} as well as reentrant surface-adsorption of proteins\cite{fries2017multivalent}
by varying the concentration of Y$^{3+}$ or other trivalent cations, (ii) pathway-controlled protein 
crystallization,\cite{zhang2011novel} (iii) clustering,\cite{zhang2012role} (iv) liquid--liquid phase 
separation,\cite{zhang2012role, zhang2012charge} and (v) lower critical solution temperature (LCST) phase 
behavior.\cite{matsarskaia2016cation} Although many aspects regarding ion--protein interactions have been qualitatively
understood, a fundamental and quantitative understanding is required for further developments in this field.\par 
Of particular interest is the LCST phase behavior for a solution of bovine serum albumin (BSA) proteins in the presence of Y$^{3+}$
ions.\cite{matsarskaia2016cation} At low temperatures, the proteins remain well dispersed in solution, whereas upon increasing
temperature up to 300 K, the proteins attract each other, and the solution separates into protein-rich and protein-poor phases. 
We note that aggregation of proteins can also be caused by thermal denaturation, but in the experiments Matsarskaia 
\textit{et al.}\cite{matsarskaia2016cation} stayed well below the protein denaturation temperature and observed LCST behavior
only for solutions containing trivalent ions.\cite{roosen2013interplay} This precludes denaturation as a mechanism
and suggests that the LCST behavior is related to ion-mediated protein aggregation.\par 
It has been suggested that the LCST behavior is due to the combination of effects associated with the solvation of the protein
and the multivalent ions, and that entropy is the driving force.\cite{matsarskaia2016cation}
However, the molecular mechanism of the LCST behavior has not been quantitatively identified.
A quantitative understanding of the thermodynamics of this process requires an accurate estimation of various entropy contributions
associated with the ion--protein complex formation and the subsequent ion-mediated protein--protein aggregation.
The total entropy change includes entropy costs due to (i) hindrance in the translation of a protein-bound ion,
(ii) restrictions on the translational and rotational motions of proteins, (iii) hydration/dehydration of the protein and ions,
and (iv) conformational changes of the protein.
The latter is mainly important for metalloregulatory allosteric proteins. Quantifying all these entropy contributions
in experiments remains a daunting task, even with the present-day techniques that provide residue-level dynamic 
information.\cite{capdevila2018functional} In this regard, molecular simulations\cite{hess2009cation, pasquier2017anomalous} 
along with accurate and robust entropy calculation techniques provide an alternative and reliable approach.\par
To understand the mechanistic details and the thermodynamic driving force for the intriguing phenomena related to ion-mediated
protein--protein interactions, we have performed large-scale molecular dynamics (MD) simulations of a single and two
BSA proteins in chloride salt solutions of Y$^{3+}$ and several other cations found in physiological conditions, such as
Na$^{+}$, Ca$^{2+}$, and Mg$^{2+}$ in the temperature range of 283--323 K. The simulation details are presented in the Methods section.
A snapshot of the initial configuration of the simulated single-protein system is shown in Figure \ref{fig:Sim1BSA}A. 
We investigate the specific nature of ion--protein interactions and quantify the free energy, various entropy contributions
as well as electrostatics of the system. Our study reveals crucial solvation/desolvation phenomena giving rise to an 
entropic driving force for ion--protein binding, in contrast to common expectations.
From simulations of the systems involving two BSA proteins, it is found that Y$^{3+}$ ions link the two proteins to form a dimer.  
Hence, the process of ion-mediated protein-protein binding is argued to be entropy-driven, as a large number of tightly bound
water molecules are released from the proteins and the mediating cations' surfaces to the bulk solution.\par
%
\section*{Results}
\subsection*{BSA Protein--Ion Interaction and Ion Binding Kinetics}
\begin{figure}
\centering
\includegraphics[width=.99\linewidth]{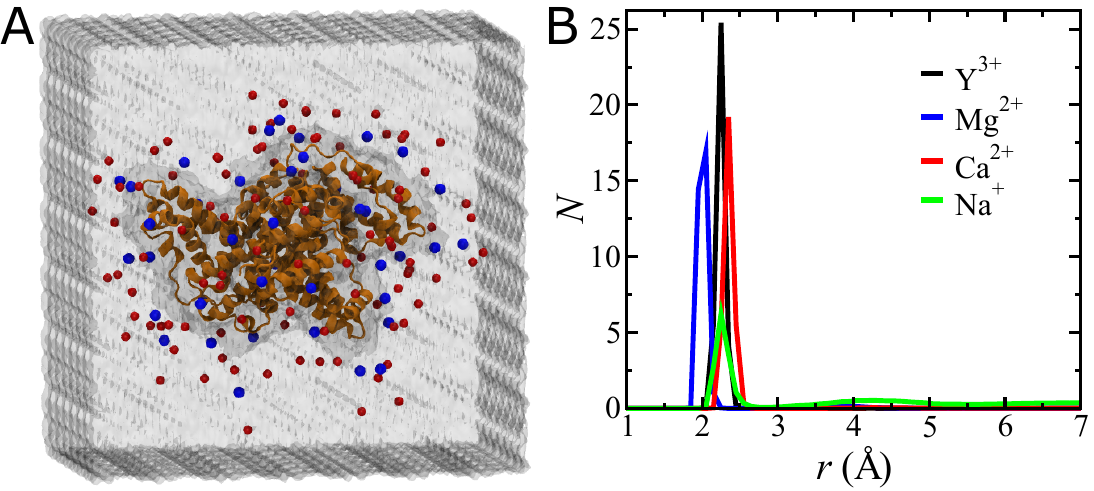}
\caption{(A) Snapshot for the starting simulation box (of size 13.2$\times$13.2$\times$13.1 nm$^{3}$) containing
a single BSA protein in 30 mM YCl$_3$ solution.
The protein (orange) is shown in cartoon representation. Y$^{3+}$ (blue) and Cl$^-$ (red) ions are represented as
VDW spheres. Water molecules are represented as a continuum in semi-transparent mode for clarity. 
Note that the system was also simulated in other salt solutions (MgCl$_2$, CaCl$_2$, and NaCl) of the same ionic
strength as in the case of YCl$_3$.
(B) For each type of cation, the total number of cations $N$ found within shells of width $dr$ $=0.1$ \AA,
present at a shortest distance $r$ from the protein surface, is plotted as a function of $r$ for the simulation performed
at 303 K. $N(r)$ for each cation type is averaged over the last 100 ns of the simulation time. Note that $N(r)$ is enhanced
near the protein surface, representing the strong affinity of cations for the negatively
charged protein ($-16$ $e$ at pH 7).
}
\label{fig:Sim1BSA}
\end{figure}

To investigate the nature of ion--protein interactions, we calculate the number distribution of ions $N(r)$ along the protein's
surface-normal direction. $N(r)$ for the cations are shown in Figure \ref{fig:Sim1BSA}B, while $N(r)$ for Cl$^-$
ion for the different ionic solutions are plotted in Figure S1 in the supporting information (SI). We find that cations are mostly
present near the protein, with the relative propensity of binding showing the following trend: monovalent < divalent < trivalent.
These cations predominantly pair with the negatively charged carboxylate groups of aspartate and glutamate surface residues of the
protein. Interestingly, even in NaCl solution, Cl$^-$ ions are found to be largely present near the protein, and the number of Cl$^-$
ions present near the protein decreases in the following order: YCl$_3$ > MgCl$_2$ $\approx$ CaCl$_2$ > NaCl (Figure S1 in the SI).
This suggests that Cl$^-$ ions interact with the --NH$_{3}^{+}$ group of the protein surface residues, and also interact, via ion-pair
formation, with the cations present in the vicinity of the protein.\par
A protein surface is, however, far from uniform and if some extended patches are present on its surface, strong affinity
of multivalent ions is expected even if the net charge of the protein is small or even of the opposite sign.\cite{yigit2015charged}
We indeed find a positively charged patch and a few extended negatively charged patches from the electrostatic potential map for
BSA (Figure S2A in the SI). We find higher density of cations (anions) near negatively (positively) charged patches even
for monovalent ions (Figure S2B,C in the SI).\par
To check how tightly the cations are bound to the protein, we monitor their binding/unbinding kinetics.
An ion is defined as bound if it is within a cutoff distance $r_c$ from any atom of the protein, otherwise the ion
is unbound or free. From the $N(r)$ plot in Figure \ref{fig:Sim1BSA}B, $r_{c}$$'s$ for the different cations are chosen as 2.8 \AA{} (Na$^{+}$),
2.7 \AA{} (Ca$^{2+}$), 2.3 \AA{} (Mg$^{2+}$), and 2.5 \AA{} (Y$^{3+}$). 
We find intermittent binding/unbinding events for both Na$^{+}$ and Ca$^{2+}$ ions (Figure S3 in the SI). While the
binding/unbinding events for Na$^{+}$ ions are frequent, prolonged bindings are observed for Ca$^{2+}$ ions. For these 
two cation types, the binding time, \textit{i.e.}, the duration for which an ion remains bound once it comes within distance of $r_c$
from the protein, is broadly distributed, owing to the surface heterogeneity of the protein. 
In contrast, only one unbinding event is observed for Mg$^{2+}$ within 1.27 $\mu$s, whereas no unbinding of Y$^{3+}$
is seen within 1.45 $\mu$s (Figure S3 in the SI). As the water escape time in the first solvation shell of Mg$^{2+}$
is $\sim$1.5 $\mu$s,\cite{lee2017ultrasensitivity} it presumably requires very long simulations (100 $\mu$s to 10 ms) to
obtain sufficient unbinding statistics for Mg$^{2+}$ and Y$^{3+}$ ions. Performing such long,
all-atom simulations for our system is out of reach of our computational capabilities.\par
%
\begin{figure}
\centering
\includegraphics[width=.99\linewidth]{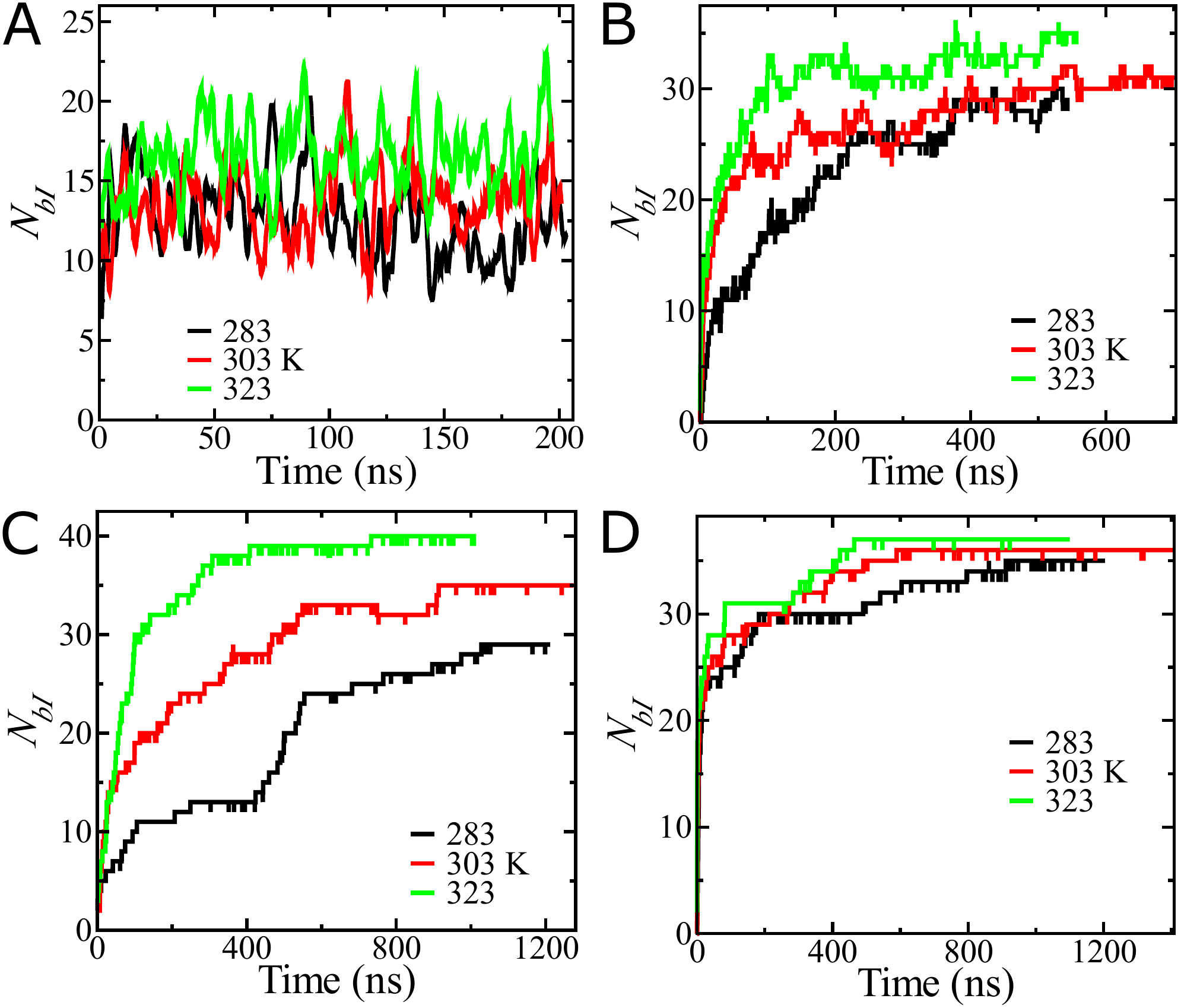}
\caption{Time series of the total number of protein-bound cations ($N_{bI}$) at several temperatures for Na$^{+}$ (A),
Ca$^{2+}$ (B), Mg$^{2+}$ (C) and Y$^{3+}$ (D) ions. For the multivalent cations,
$N_{bI}$ increases upon increasing temperature.
}
\label{fig:BindKin}
\end{figure}
For each cation type, the total number of protein-bound cations, $N_{bI}$, is plotted as a function of the simulation time at three different 
temperatures in Figure \ref{fig:BindKin}. No ion is bound to the protein at the beginning of a simulation, and $N_{bI}$ gradually increases with 
the simulation time. $N_{bI}$ eventually reaches a saturation value, at a time required for equilibration of the ion distribution
around the protein. This ion equilibration time differs for each cation, which can be rationalized by considering the ion--water exchange
kinetics that strongly depends on the cation's charge and size.\cite{lee2017ultrasensitivity} Counterintuitively, we find from 
Figure \ref{fig:BindKin} that $N_{bI}$ increases with increasing temperature. This effect is prominent for all the cations, except Na$^+$. 
In contrast,  the number of protein-bound water, \textit{i.e.}, the total number of water molecules present within 3 \AA{} from the protein
surface decreases with the increase in temperature as expected (Figure S4 in the SI). Although an increase in the binding affinity 
of any two objects by raising the temperature is not new---\textit{e.g.}, hydrophobic interaction strength increases with 
temperature,\cite{chandler2005interfaces} it is surprising to be observed in a system involving strong electrostatic
interactions and can be rationalized by the temperature dependence of dielectric and hydration 
effects.\cite{israelachvili2015intermolecular, sedlmeier2013solvation}
For a quantitative understanding of this, we calculate various thermodynamic quantities such as the free energy, enthalpy,
and various entropy contributions as discussed below.\par    
%
%
\subsection*{Thermodynamics of Cation Binding to the Protein}
\begin{figure}
\centering
\includegraphics[width=.99\linewidth]{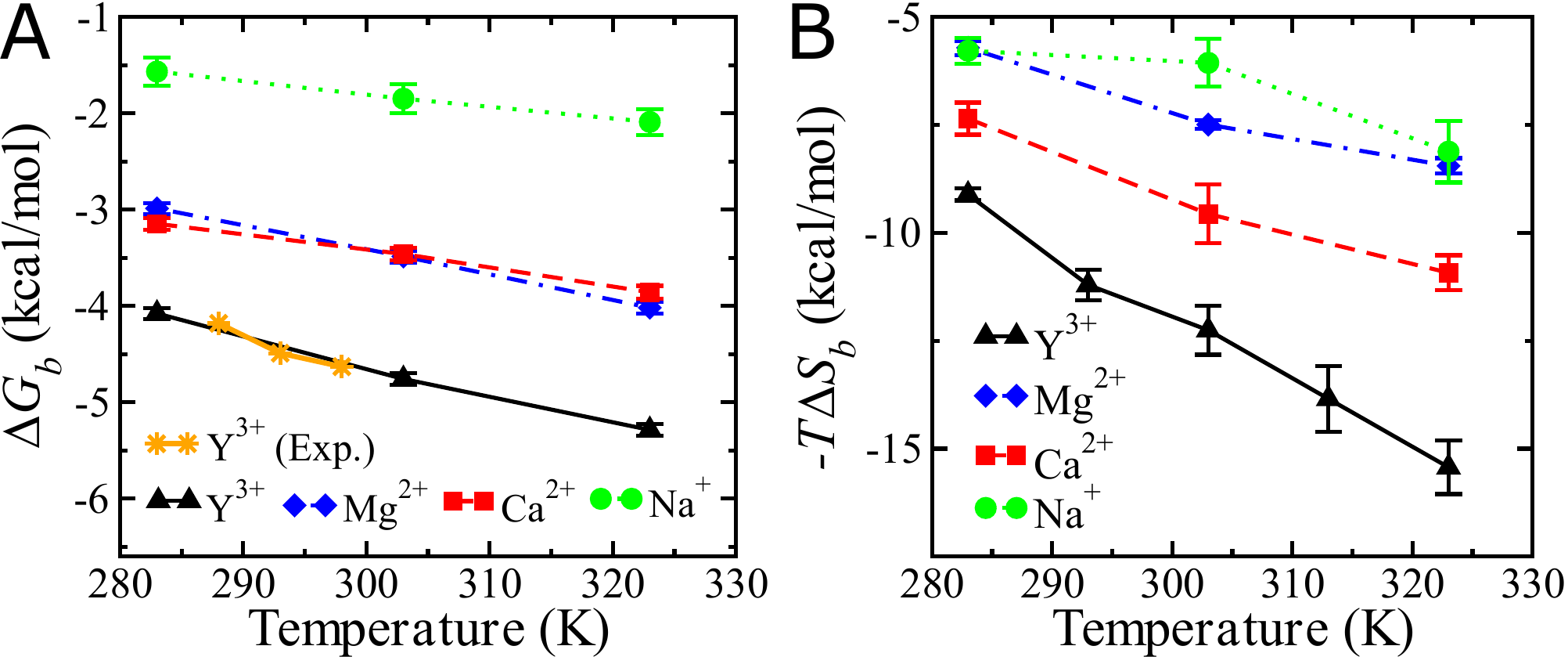}
\caption{Temperature dependence of (A) the free energy, $\Delta G_b$, and (B) the total entropy contribution, $-T\Delta S_b$, 
for each cation binding to the protein. Error bars represent the standard deviation. The different lines are for guiding the eye.
The experimental binding free energies for Y$^{3+}$ at different temperatures shown in (A) are taken from Ref.\cite{matsarskaia2016cation} 
}
\label{fig:BindTher}
\end{figure}

The free energy of a cation binding to the protein, $\Delta G_b$, for temperatures in the range of 283--323 K is shown in Figure \ref{fig:BindTher}A
(see Methods for the calculation details). For each cation type, $\Delta G_b$ is always negative, and its magnitude increases
with the increase in temperature. $|\Delta G_b|$ follows the trend: Na$^+$ < Ca$^{2+}$ $\approx$ Mg$^{2+}$ < Y$^{3+}$.
By changing temperature from 283 K to 323 K, we see the highest change in $\Delta G_b$
for Y$^{3+}$ ($-1.21$ kcal/mol), whereas the least change is observed for Na$^+$ binding ($-0.52$ kcal/mol). The changes in
$\Delta G_b$ for Ca$^{2+}$ and Mg$^{2+}$ ions are $-0.71$ and $-1.03$ kcal/mol, respectively.
\par
The increase in binding affinity of the cations with solely increasing temperature (Figure \ref{fig:BindTher}A) cannot be explained
by considering the energy of binding, for purely thermodynamic reasons, as described in the SI, section 1. 
Further, it should be noted that since the dielectric constant of water decreases as $\sim T^{-3/2}$, any electrostatic interaction in
water is predominantly entropic in nature.\cite{israelachvili2015intermolecular, sedlmeier2013solvation} Therefore, entropy must
be playing a dominant role here.\par
\begin{figure}
\centering
\includegraphics[width=.99\linewidth]{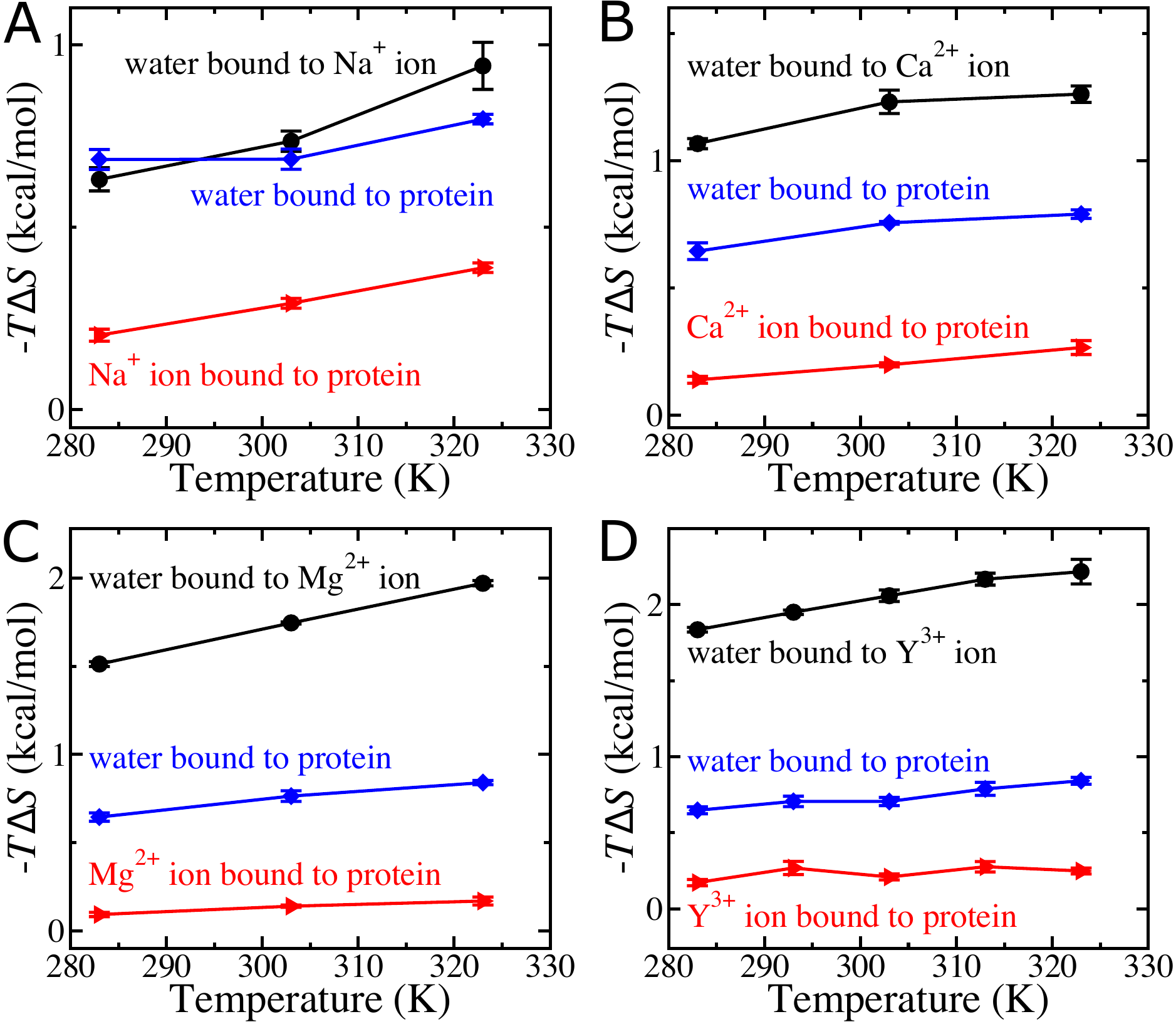}
\caption{Entropy contributions $\Delta S_{P,I}$ (red), $\Delta S_{I,W}$ (black), $\Delta S_{P,W}$ (blue) from Eq. \ref{eq:TotEntropy}
as a function of temperature for the protein in NaCl (A), CaCl$_2$ (B), MgCl$_2$ (C), and YCl$_3$ (D) solutions. Error bars represent
the standard deviation. The lines are for guiding the eye.
}
\label{fig:Entropy}
\end{figure}

The binding free energy for an ion is given by
\begin{equation}\label{eq:ThermoRel}
\Delta G_{b}(T)=\Delta E_{b}(T)-T\Delta S_{b}(T),
\end{equation}
where $\Delta E_{b}$ and $\Delta S_{b}$ are the energy and entropy of binding, respectively and $T$ is the temperature. 
For the calculation of $\Delta S_{b}$, one needs to correctly account for
``hydration effects'' associated with the ion binding process, such as partial desolvation of both the protein and ion.
The radial distribution functions for water molecules around a cation, both free in solution and bound to the protein surface,
clearly show partial dehydration of the first and second solvation shells (SS$'s$) of each cation (Figure S5 in the SI).
$\Delta S_{b}$ in Eq. \ref{eq:ThermoRel} consists of three terms---the loss in entropy of a protein-bound ion ($\Delta S_{P,I}$),
the gain in entropy due to release of tightly-bound water molecules from the first and second SS$'s$ of the ion ($\Delta S_{I,W}$),
and the gain in entropy of water molecules released to the bulk due to desolvation of the protein surface residue where the ion binds
($\Delta S_{P,W}$). Together, it can be written as 
\begin{equation}\label{eq:TotEntropy}
\Delta S_{b}=\Delta S_{P,I}-\Delta S_{I,W}-\Delta S_{P,W}.
\end{equation}
We have used the two-phase thermodynamic (2PT) method\cite{lin2003two, lin2010two, pascal2011thermodynamics}
to calculate all the terms on the right hand side of Eq. \ref{eq:TotEntropy}.
The theory of the 2PT method is described in the SI, section 2 and calculation details are given in the Methods section. 
We first validate the 2PT method for ionic solutions by reproducing from the simulation data the experimental ion
hydration entropy ($\Delta S_{hyd}$) for the different ion types (see Table S1 in the SI). 
Then, we proceed, using 2PT, with calculations of the
entropy differences for a protein-bound ion, a protein-bound water, and a water in the first SS of the cation as shown
in Figure \ref{fig:Entropy}, as well as the entropy difference of a water in the second SS of the cation as shown in Figure S6 in the SI.
Note that for calculations of the various entropy contributions shown in Figures \ref{fig:Entropy} and S6, the reference values are taken
as the respective absolute entropies in the bulk water. $\Delta S_{I,W}$ in Eq. \ref{eq:TotEntropy} is then calculated by multiplying
the per water entropy differences with the corresponding numbers of water molecules released in the partial dehydration of both the first
and second SS$'s$ of the cation (values given in Table S2 in the SI), and adding both terms. Similarly,
$\Delta S_{P,W}$ in Eq. \ref{eq:TotEntropy} is evaluated by multiplying the per water entropy difference with the number of water molecules
released in the partial dehydration of the protein surface residue (values given in Table S2 in the SI). From Figure \ref{fig:Entropy}, we see
that the entropy loss of a protein-bound cation is more than compensated by the entropy gain of water molecules released to bulk by
the partial dehydration of both the cation and protein. The cation desolvation entropy contributes the highest to the thermodynamics
of protein--ion binding for all the multivalent cations, whereas both the protein and ion desolvation entropies contribute equally for
Na$^+$ binding.\par
The total entropy contribution ($-T\Delta S_{b}$ in Eq. \ref{eq:ThermoRel}) for each cation plotted in Figure \ref{fig:BindTher}B is 
always negative, and it decreases (becomes more negative) with increasing temperature. 
We have also estimated $-T\Delta S_b$ from the temperature dependence of $\Delta G_b$ using the thermodynamic
relation $\Delta S_{b}=-\partial \Delta G_{b}/\partial T$, and we find that the temperature dependence trend is the same as obtained
from the 2PT method, though the values obtained from both methods match only semi-quantitatively (see Figure S7 in the SI for comparison).
For each cation, $-T\Delta S_{b}$ is more negative than the binding free energy $\Delta G_{b}$ throughout the temperature range
studied in this work (Figure \ref{fig:BindTher}). Therefore, the process of a cation binding to the protein is entropy driven. The above
observations, in particular, explain the enhancement of the protein-binding affinity of a multivalent cation with increasing temperature
(Figure \ref{fig:BindKin}). The total entropy contribution as shown in Figure \ref{fig:BindTher}B is the highest for Y$^{3+}$
across the whole temperature range, followed by Ca$^{2+}$ > Mg$^{2+}$ $\approx$ Na$^+$---representing a delicate dependency of
entropy on the charge and size of a cation. Note that though the entropy contribution due to a water molecule released from
Mg$^{2+}$ is more than that for Na$^+$ and Ca$^{2+}$ ions (Figure \ref{fig:Entropy}), the altered trend in $\Delta S_{b}$ for Mg$^{2+}$
in Figure \ref{fig:BindTher}B is rationalized by the lower number of water molecules released in the process of a Mg$^{2+}$ ion binding,
compared to that for Na$^+$ and Ca$^{2+}$ bindings.\par
The large (and negative) value of the entropy contribution, $-T\Delta S_{b}$, must be partially compensated by a positive binding
energy $\Delta E_{b}$ to result in a small (and negative) value of the binding free energy $\Delta G_{b}$. $\Delta E_{b}$, calculated by 
using the thermodynamic relation given in Eq. \ref{eq:ThermoRel}, is plotted as a function of temperature in Figure S8 in the SI. 
$\Delta E_{b}$ is positive throughout the temperature range, in agreement with the experiment,\cite{matsarskaia2016cation}
but is comparable to the magnitude of $-T\Delta S_{b}$. 
The increase in $\Delta E_{b}$ with temperature (Figure S8) can be rationalized by the enhancement in the electrostatic
interaction strength due to the decrease in the water dielectric constant $\epsilon$, as explained below.\cite{israelachvili2015intermolecular}
The electrostatic free energy $\Delta G \propto \epsilon^{-1}$ and $\epsilon \propto T^{-\alpha}$, thus $\Delta G=-CT^{\alpha}$.
Here $C$ is a constant and the negative sign is due to $\Delta G <0$ in our case. 
The entropy follows as $\Delta S = -\partial\Delta G / \partial T = C\alpha T^{\alpha-1}$.
The internal energy results as $\Delta E = \Delta G + T\Delta S = -CT^{\alpha} + C\alpha T^{\alpha} = C (\alpha -1) T^{\alpha}$.
As long as the exponent $\alpha>1$, $\Delta E$ is always positive and increases as $T^{\alpha}$. 
For pure water $\alpha>1$ at all temperatures (Figure S9 in the SI). Although $\alpha$ slightly decreases with the addition of salt 
(\textit{viz.} 1 M NaCl solution in Figure S9), $\alpha$ is significantly greater than 1 for the temperature regime investigated
in our simulations, which explains the observed increase in $\Delta E_{b}$ with increasing temperature.\par
The temperature-dependent increase in $\Delta E_{b}$ follows the trend: 
Y$^{3+}$ > Ca$^{2+}$ > Na$^+$ $\approx$ Mg$^{2+}$ (Figure S8 in the SI). By changing temperature from 283 K to 323 K, the change in
$\Delta E_{b}$ for Na$^+$, Ca$^{2+}$, Mg$^{2+}$, and Y$^{3+}$ is found to be 1.82, 2.86, 1.71, and 5.11 kcal/mol, respectively.\par
The large value of $\Delta E_{b}$ can be understood by considering the energetic penalties associated with the desolvation of both
the protein and cation. For example, $\Delta E_{b}$ for Y$^{3+}$ ion at 300 K decreases from 7.50 to 3.71 kcal/mol if we
exclude the contribution due to the dehydration of the second SS of Y$^{3+}$ (Figure S10B in the SI).  Figure S10 also highlights that
the effect of the second SS is significant for the accurate description of solvation thermodynamics of cations, and
cannot be neglected even for monovalent ions, \textit{e.g.}, Na$^{+}$.\par
%
%
\subsection*{Preferential Interaction Coefficients}
The interaction of ions with proteins, whether these are enriched or depleted from the protein surface, can be
quantified by experimentally measuring the preferential interaction coefficient $\Gamma_{23}$. The thermodynamic
definition of $\Gamma_{23}$ is the change in chemical potential of the protein due to the addition of ions;
it can also be expressed as the change in ion concentration to maintain constant chemical potential when a protein
is added to the solution:\cite{pierce2008recent}
\begin{equation}\label{eq:PIC1}
\Gamma_{23}=-\left(\frac{\partial\mu_2}{\partial\mu_3}\right)_{m_2,T,P}
           =-\left(\frac{\partial m_3}{\partial m_2}\right)_{\mu_3,T,P},
\end{equation}
where $\mu$ is the chemical potential, $m$ is the molal concentration, and the subscripts 1, 2, and 3 stand for water,
protein, and ion, respectively. Record \textit{et al.},\cite{record1995interpretation} based on the molal concentration
definition, developed a two-domain molecular model for the estimation of $\Gamma_{23}$ in terms of the difference
in ion concentration in the local domain near the protein surface and the bulk solution as follows:
\begin{equation}\label{eq:PIC2}
\Gamma_{23}=\bigg\langle N_3^{local} - N_1^{local}\bigg[\frac{N_3^{bulk}}{N_1^{bulk}}\bigg] \bigg\rangle,
\end{equation}
where $N_i$ is the number of molecules of type $i$ and $\langle \cdot \rangle$ represents the time average. 
For the calculation of $\Gamma_{23}$ using Eq. \ref{eq:PIC2} a boundary or a distance cutoff needs to be chosen
for defining the local and bulk domain, but the choice is arbitrary.   
$\Gamma_{23}$ is instead estimated at each value of $r$, the distance from the protein surface, asuming that it is
the boundary: $\Gamma_{23}(r) = \langle N_3(r) - N_1(r)[N_3^{bulk}/N_1^{bulk}] \rangle$. 
The distance $r^*$ after which $\Gamma_{23}(r)$ becomes constant is defined as the actual boundary. 
In our simulations the total numbers of water molecules ($N_1$) and
ions ($N_3$) are constant, thus the above expression for $\Gamma_{23}(r)$ is further simplified as:\cite{shukla2009molecular}
\begin{equation}\label{eq:PIC3}
\Gamma_{23}(r)=\bigg\langle N_3(r) - N_1(r)\bigg[\frac{N_3-N_3(r)}{N_1-N_1(r)}\bigg] \bigg\rangle.
\end{equation}
\par
\begin{figure}
\centering
\includegraphics[width=.65\linewidth]{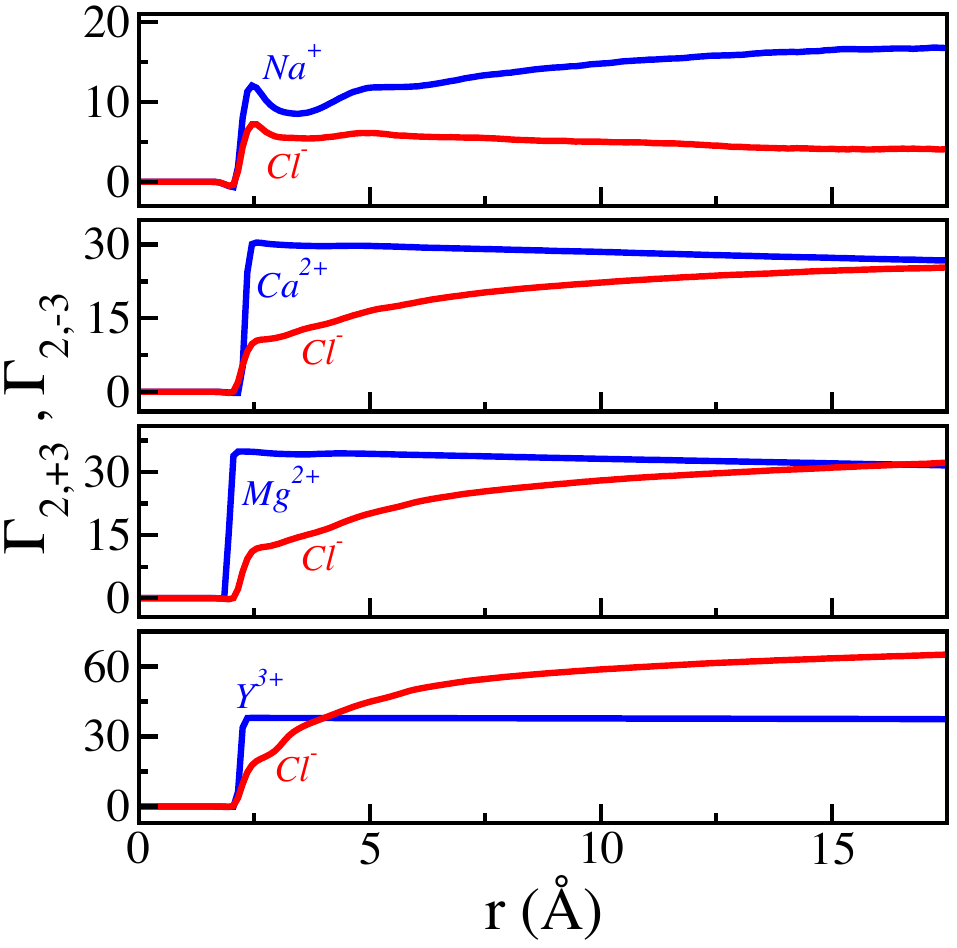}
\caption{Preferential interaction coefficients for cations $\Gamma_{2,+3}$ and anions $\Gamma_{2,-3}$ of the different
salt solutions at 303 K as a function of the boundary distance cutoff $r$. $\Gamma_{2,+3}(r)$ reaches a saturation value rather quickly
compared to $\Gamma_{2,-3}(r)$, for solutions containing multivalent cations.
}
\label{fig:PIC}
\end{figure}

In a salt solution, cations and anions are distributed around the protein. We obtain preferential interaction parameters
for the cation $\Gamma_{2,+3}(r)$ and anion $\Gamma_{2,-3}(r)$ separately by using $N_{+/-3}(r)$ as the cation or anion
distribution, respectively in Eq. \ref{eq:PIC3}. $\Gamma_{2,+3}(r)$ and $\Gamma_{2,-3}(r)$ are shown for different salt solutions
in Figure \ref{fig:PIC}. Experimentally, it is impossible, however, to separate the cationic and anionic contribution to the
measured value of  $\Gamma_{23}$ for a salt solution. For a salt of monovalent cation and anion, the preferential
interaction parameter is given by\cite{record1995interpretation} 
\begin{equation}\label{eq:PIC4}
\Gamma_{23}=\frac{1}{2}\big( \Gamma_{2,+3}+\Gamma_{2,-3}-|Q_2| \big),
\end{equation}
where $|Q_2|$ is the protein’s net charge that is subtracted from $\Gamma_{23}$, as $Q_2$ counterions
(cations in case of BSA protein) are accumulated near the protein surface to neutralize its charge and do not contribute
to the preferential interaction. For a salt of multivalent cation/anion, it is straight-forward to generalize Eq. \ref{eq:PIC4}
by scaling $\Gamma_{2,+3}(r)$, $\Gamma_{2,-3}(r)$, and $Q_2$ with valency of the anion $z_-$, valency of the cation $z_+$,
and charge on the counterion $z_+$, respectively.
\begin{equation}\label{eq:PIC5}
	\Gamma_{23}=\frac{1}{2}\bigg( \frac{\Gamma_{2,+3}}{z_-}+\frac{\Gamma_{2,-3}}{z_+}-\frac{|Q_2|}{z_+} \bigg).
\end{equation}
\par
For the BSA protein, using in Eq. \ref{eq:PIC5}  $\Gamma_{2,+3}$ and $\Gamma_{2,-3}$ at $r^*=17$ \AA{} (by which all curves
reach their respective saturation values as seen in Figure \ref{fig:PIC}), we obtain preferential
interaction coefficients for different salts: NaCl ($\Gamma_{23}=2.44$), CaCl$_2$ ($\Gamma_{23}=15.67$), MgCl$_2$
($\Gamma_{23}=19.83$), and YCl$_3$ ($\Gamma_{23}=26.87$). Positive values of $\Gamma_{23}$ for all the different
salt types reflect that these salt ions are attracted towards the protein surface. For salt containing multivalent ions,
$\Gamma_{23}$ is significantly larger than that for NaCl, which suggests that addition of trivalent ions in the protein
solution affects the solution stability\cite{zhang2010universality} and stabilizes protein dimer formation as seen
in our simulations.\par  
%
%
\subsection*{$\zeta$-Potential of the Protein and the Protein--Protein Interaction}
$\zeta$-potential measurements for a protein in an ionic solution report on charge compensation by the counterions and 
thus have direct implications for protein--protein association and the phase behavior of the solution. $\zeta$-potentials are
defined by the electrophoretic mobility.\cite{roosen2013interplay, matsarskaia2016cation, kubivckova2012overcharging}
From the simulation data, we calculate the surface potential at one ionic diameter away from the protein surface (see Methods).
Note that the surface potential typically serves as a good approximation for the $\zeta$-potential for proteins and
colloidal systems; however, the surface and $\zeta$-potential values might differ significantly for extended
surfaces with high surface charge densities.\cite{uematsu2018analytical}\par
\begin{figure}
\centering
\includegraphics[width=.60\linewidth]{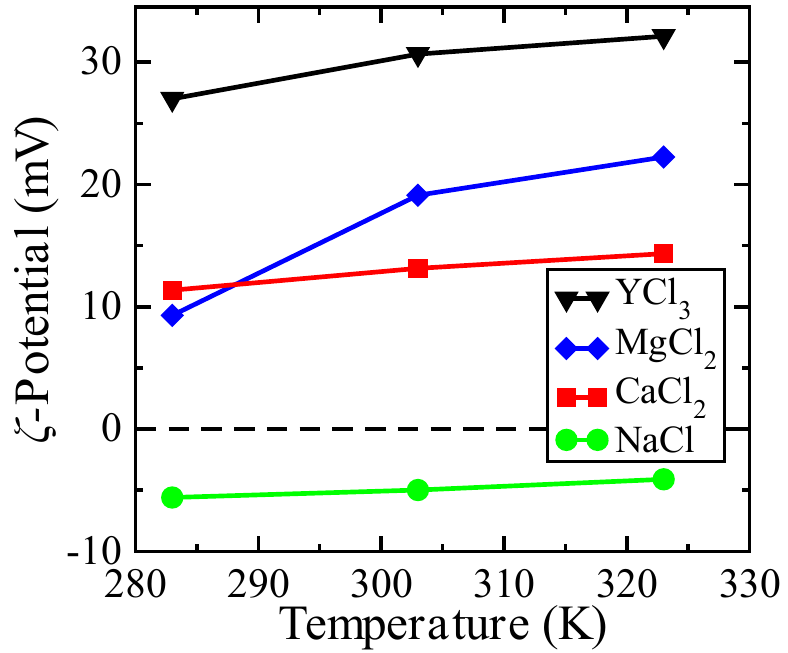}
\caption{Temperature dependence of the $\zeta$-potential of the protein in various salt solutions. The lines are guide to the eye only.
}
\label{fig:ZetaPot}
\end{figure}

   
As shown in Figure \ref{fig:ZetaPot}, the $\zeta$-potential of the protein in the NaCl solution is negative at all temperatures,
as expected based on the protein net charge of $-16$ $e$. 
In contrast, the $\zeta$-potential is positive for all multivalent 
cation-chloride solutions at all temperatures, indicating sign reversal of the effective charge of the protein (Figure \ref{fig:ZetaPot}).
This \textit{charge inversion} phenomenon in the presence of multivalent cations is due to strong interactions of the cations
predominantly with the COO$^-$ groups of the protein's surface residues and can be rationalized by considering strong charge-charge
correlations.\cite{grosberg2002colloquium} Note that similar charge reversal of proteins in the presence of trivalent cations
has also been reported both in experiments\cite{zhang2008reentrant, kubivckova2012overcharging, matsarskaia2016cation}
and simulations\cite{pasquier2017anomalous} as well as in a coarse-grained analytical model.\cite{roosen2014ion}
As shown in Figure \ref{fig:ZetaPot}, with the increase in temperature, the $\zeta$-potential of the protein increases for all cation types;
the highest change is seen for MgCl$_2$, whereas the effect is minimal for NaCl. The $\zeta$-potential of the protein at 283 K
is higher in CaCl$_2$ solution than in MgCl$_2$ solution, and vice versa at 323 K. These observations are consistent with the
trends for the temperature dependence of binding free energies of the different cations (Figure \ref{fig:BindTher}A).\par
%
%
\subsection*{Protein--Protein Binding Mediated by Cation Bridges}
\begin{figure*}
\centering
\includegraphics[width=.99\linewidth]{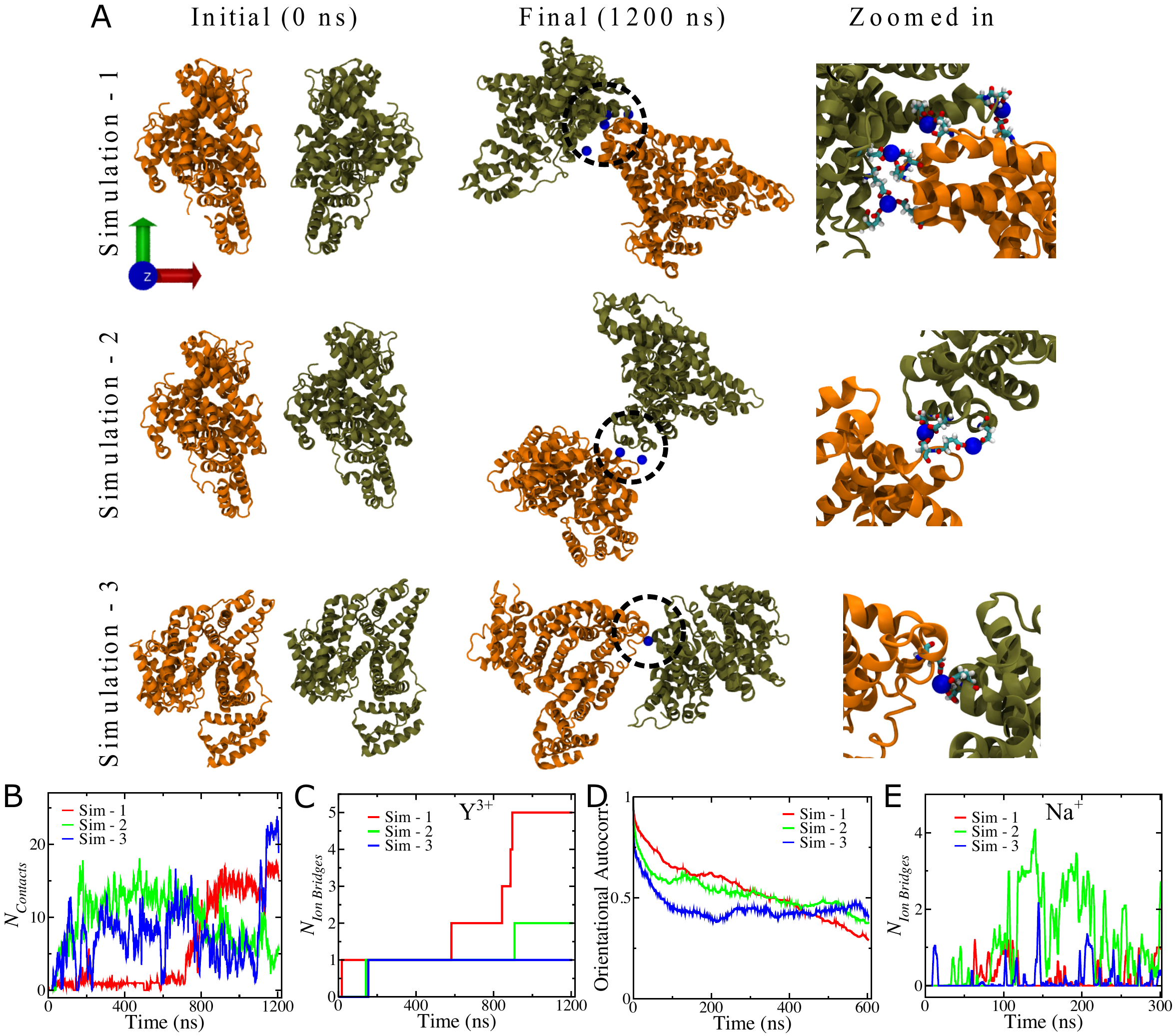}
\caption{Results for two BSA proteins (A--D) in 30 mM YCl$_{3}$ and (E) 180 mM Nacl solution at 303 K. (A) The initial and final configuration
of two BSA proteins (represented in orange and tan). Bridging Y$^{3+}$ ions in the final structure are shown as blue spheres,
and amino acid residues of the two proteins involved in the formation of ion bridges are highlighted in the ball--stick representation
(see the right panel for the zoomed-in version). Water and other ions are omitted for clarity. (B) Time series of the total
number of inter protein residue–residue contacts. (C) Time series of the total number of Y$^{3+}$ ion bridges that link
the two proteins to form a dimer. (D) The average autocorrelation function for the relative orientation angles between
the principal axes of two proteins as a function of time delay. (E) Time series of the total number of Na$^{+}$ ion bridges
that form transiently between two BSA proteins in 180 mM NaCl solution at 303 K. All observables are defined in Methods. 
}
\label{fig:Sim2BSA}
\end{figure*}

Protein aggregation seen in experiments\cite{matsarskaia2016cation} was hypothesized to be mediated by cation bridges.\cite{zhang2011novel}
To explicitly demonstrate the multivalent ion-mediated protein--protein binding, we have performed three independent simulations with
different orientations of two BSA proteins in YCl$_{3}$ solution, as shown in Figure \ref{fig:Sim2BSA}A (left panel). In every simulation, 
we find that two BSA proteins, which are initially placed far apart, approach each other (see the timeseries of the total number
of inter-protein residue--residue contacts in Figure \ref{fig:Sim2BSA}B) and eventually form a dimer mediated by 1--5 Y$^{3+}$ ions (see snapshots
in Figure \ref{fig:Sim2BSA}A [middle panel]). The Y$^{3+}$ ion bridges remain stable over a 1 $\mu$s timescale, as evident from the time series plot
for the number of bridging cations (Figure \ref{fig:Sim2BSA}C). A Y$^{3+}$ ion bridge is stabilized by coordination of multiple caboxylate groups
of each protein with the cation, as evident from snapshots in Figure \ref{fig:Sim2BSA}A (right panel). 
Note that even for the stable, Y$^{3+}$ ion-bridged protein dimer complex, the relative orientation between the two proteins changes
over time but very slowly (see the orientational autocorrelation function in Figure \ref{fig:Sim2BSA}D). This reveals the conformational
flexibility of the protein dimer complex.\par
To compare monovalent and multivalent cations, we have also simulated the above three systems (shown in Figure \ref{fig:Sim2BSA}A) in NaCl
solution, at the equivalent ionic strength as for YCl$_{3}$ solution. In sharp contrast to the case of YCl$_{3}$,
we find that Na$^{+}$ ion bridges between two BSA proteins form transiently and remain stable only for 1--20 ns (Figure \ref{fig:Sim2BSA}E).
These results demonstrate the need for multivalent ions in protein cluster formations, in agreement with experiments.\cite{zhang2008reentrant}\par 
%
\section*{Discussion}
The temperature behavior of the $\zeta$-potential found in Figure \ref{fig:ZetaPot} is in qualitative agreement with the
experiments.\cite{matsarskaia2016cation} 
As the $\zeta$-potential is influenced by the number of surface-bound ions and the binding affinity of ions increases with temperature,
the $\zeta$-potential is expected to increase with temperature irrespective of the salt concentration of the solution.
$\zeta$-potential values estimated from our simulations, however, are larger than that reported in the experiments\cite{matsarskaia2016cation}
presumably because of the YCl$_{3}$ concentration difference.
The $\zeta$-potential increases with increasing YCl$_{3}$ concentration as found in experiments,\cite{matsarskaia2016cation}
thus we expect the simulation and experimental results to match if the same salt concentration is used.\par 
It should be noted that the YCl$_3$ concentration used in our simulations is 30 mM which is much higher than the 1 mM
concentration used in the experiment. 
A direct comparison between all-atom simulations and the experiments\cite{matsarskaia2016cation} at low concentration
of multivalent ions is rather difficult for the following reason. We consider a higher YCl$_3$ ion concentration in our 
simulations in order to obtain statistically converged results with sufficient number of ions. Obtaining well-converged results for
proteins at low salt concentrations with enough number of ions would require significantly larger system sizes. Simulating such
large systems is very demanding at the all-atom level, but it is feasible at a coarse-grained level as shown in a recent
study.\cite{pasquier2017anomalous} However, the solvation effects, which are crucial for the accurate prediction of protein--ion
binding thermodynamics, are not properly taken into account in such coarse-grained simulations.\par
The LCST phase behavior found in experiments\cite{matsarskaia2016cation} can be rationalized by the temperature dependence
of the $\zeta$-potential and the microscopic picture emerges from our simulations.
For sufficiently low Y$^{3+}$ concentration, at low temperatures due to the reduced binding affinity of counterions,
the $\zeta$-potential is expected to become negative (and large) and the proteins are expected to repel each other,
keeping the solution stable. With increasing temperature, counterion binding affinity for the protein increases, and hence the
$\zeta$-potential increases and becomes positive at a sufficiently high temperature. In the temperature range (293--313 K) where
the $\zeta$-potential is small ($-$5 to $+$5 mV),\cite{matsarskaia2016cation} the proteins are predicted to attract each other,
eventually causing the solution to phase separate into protein-rich and protein-poor phases.\par
The protein--protein binding at a low concentration of multivalent ions occurs via cation bridging, as shown in Figure \ref{fig:Sim2BSA}, 
as well as suggested from experiments.\cite{zhang2011novel, matsarskaia2016cation} 
A cation bridge formation---like the first step of a cation binding to the
protein---requires desolvation of both the protein-bound cation and the surface residue of another protein that will bind to
the cation. These processes involve the release of many tightly-bound water molecules to the bulk that results in a significant
entropy gain, which contributes at least 10--15 kcal/mol (depending on the temperature) to the total free energy, as shown in
Figure \ref{fig:BindTher}B for a Y$^{3+}$ ion binding. As multiple cation bridges are formed in a protein--protein binding
(see Figure \ref{fig:Sim2BSA}, and also
found in experiments\cite{zhang2011novel}), the net entropy gain due to cation and protein desolvation more than compensates
for the translational and rotational entropy losses of the proteins during protein--protein binding. Therefore, the LCST phase
behavior\cite{matsarskaia2016cation} is entropy-driven.\par
%
%
\section*{Conclusions}
In summary, by performing fully atomistic MD simulations of a BSA protein in different cation-chloride solutions
(NaCl, CaCl$_2$, MgCl$_2$, and YCl$_3$) and by calculating various entropy contributions, we demonstrate that 
multivalent cation binding to the protein is an entropy-driven phenomenon. The loss in
entropy of a protein-bound cation is more than compensated by the entropy gain of water molecules due to the partial dehydration of
both the cation and the cation-bound surface residue of the protein. 
A particularly interesting observation is the significant difference in the binding/unbinding kinetics of Ca$^{2+}$ and Mg$^{2+}$ ions
(see Figure S3)---although having comparable binding free energies (Figure \ref{fig:BindTher}A), which can be related to the recent finding 
that the ion--water exchange kinetics strongly depends on the size of a cation.\cite{lee2017ultrasensitivity} It will thus be interesting
to investigate in future simulation studies the universality of the ion size dependence of ion--protein binding kinetics and thermodynamics.\par 
%
The $\zeta$-potential calculation shows \textit{charge inversion} of the protein in all solutions containing multivalent cations,
but not in the monovalent NaCl solution (Figure \ref{fig:ZetaPot}). 
The LCST phase behavior observed in the experiment\cite{matsarskaia2016cation} can be rationalized by considering the
temperature-dependent increase in the $\zeta$-potential of the protein and the associated \textit{charge inversion} phenomenon.
The protein--protein interaction involves: (i) the ion
binding to the protein, and (ii) the subsequent protein--protein binding by cation bridging (Figure \ref{fig:Sim2BSA}). 
In both processes many tightly-bound water molecules are released to the bulk, which results in a thermodynamic driving force
for the LCST behavior that is entropic in nature, in agreement with the experiment.\cite{matsarskaia2016cation}\par
This work shows that similarly to hydrophobic association, entropy plays a pivotal role in systems involving
strong electrostatic interactions, revealing intriguing hydration and dielectric effects.
Our results are important for the basic understanding of ion effects in soft matter and biology, and the insights gained here
will be useful in studies of ion-mediated surface adsorption and crystallization of proteins.
Moreover, molecular-level understanding of interactions of heavy metals---usually not found in healthy cells---with
different biomolecules, as studied here, can provide insights for carcinogenicity and neurotoxicity induced by exposure
to such environmental contaminants.\par
%
%
\section*{Methods}
\subsection*{Model Building and Force Field Parameters}
The initial structure of BSA protein was obtained from the crystal structure available in the protein data
bank (PDB ID: 3V03). The charge or protonation state of each residue of the protein was chosen at neutral pH 7 depending
on the residue's pK$_{a}$ value, and the assigned charges were fixed over the simulation time.
Note, however, that pK$_{a}$ depends on the ionic strength (through the activity coefficients), and the reported pK$_{a}$ values of
amino acids are typically determined in a solution of high ionic strength.\cite{reijenga2013development} In particular, the apparent
pK$_{a}$ values of carboxyl groups shift up slightly in the presence of multivalent cations at low salt concentrations. If
the pH of the solution differs from 7 in an experiment due to the CO$_2$ content in air (which lowers the water pH down to
5.6 \cite{uematsu2017charged}), this could make some of the acidic peptide groups less charged. But, the experiments described
in Matsarskaia \textit{et al.}\cite{matsarskaia2016cation} were performed in air and in ultrapure (MilliQ, 18.2 Mega Ohm) water
which had previously been degassed under vacuum to eliminate the CO$_2$ contributions. Also, it is known from experiments that the
addition of multivalent metal cations such as Al$^{3+}$ and Fe$^{3+}$ changes the pH of the solution due to hydrolysis of these
cations, which can change the charge states of the protein surface residues. However, this effect is less significant
for Y$^{3+}$ ion.\cite{roosen2013interplay}
Therefore, our assigned fixed charges of the protein residues at pH 7 for the different cations (Na$^+$, Ca$^{2+}$, Mg$^{2+}$,
and Y$^{3+}$) is assumed to mimic the experimental conditions sufficiently well.\par
The ff14SB force field parameters\cite{maier2015ff14sb} were used for the protein. The system was solvated with 
TIP3P\cite{jorgensen1983comparison} water model using the \textit{xleap} module of the AMBER17 tools\cite{amber17tools} in a way
such that there exists at least 17 \AA{} solvation shell in between the solute and simulation box wall. The final unit shell for
simulation is a rectangular box of size 13.2$\times$13.2$\times$13.1 nm$^{3}$ that contains $\sim$200,000 atoms (Figure \ref{fig:Sim1BSA}A). 
The system was simulated in four different salt solutions, namely NaCl, MgCl$_{2}$, CaCl$_{2}$ and YCl$_{3}$.
Depending on the ion type, an appropriate number of counterions were added to ensure the charge neutrality of the
simulation unit shell. To simulate the system at a specified salt concentration, enough numbers of counterions/coions, estimated
from the mole fraction of counterions/coions and water, were further added to the system. For YCl$_{3}$, the system was simulated
at 30 mM salt solution. For the other salts, the system was simulated at the equivalent ionic strength as in the case of YCl$_{3}$,
e.g., 180 mM for NaCl.
Especially for multivalent ions, the electronic polarization effect contributes significantly to the total interaction energy of such
an ion with another charged object. The recently developed Li/Merz ion
parameters\cite{li2013taking, li2014parameterization, li2015systematic} with 12-6-4 Lennard-Jones (LJ)-type nonbonded interaction
terms take care of the electronic polarization effect and have been shown to well reproduce the experimental measurables, such as the
ion--oxygen (of water) distance, the ion--water coordination number, and the hydration free energy of mono- and multi-valent ions.
We have provided in Table S1 in the SI the structural parameters and entropy of ion hydration for the different ions
calculated from our simulation data, which quantitatively match with the corresponding experimental values.
Therefore, we used Li/Merz ion parameters for an accurate modeling of the ion--water and ion--protein interactions.\par
\subsection*{MD Simulation Details}
All the simulations were performed using the PMEMD module of the AMBER14 package.\cite{case2005amber} Periodic boundary condition
was used for all the simulations. Bonds involving hydrogen atoms were constrained using the SHAKE 
algorithm\cite{ryckaert1977numerical} that allowed the use of a time step of 2 fs for the integration of Newton's equation
of motion. The temperature of the system was maintained using a Langevin thermostat\cite{van1988leap} with the collision
frequency of 5.0 ps$^{-1}$.
Berendsen weak coupling method\cite{berendsen1984molecular} was used to apply a pressure of 1 atm with isotropic position
scaling with a pressure relaxation time constant of 2.0 ps. Particle mesh Ewald\cite{darden1993particle} sum was used
to compute long-range electrostatic interactions with a real space cutoff of 10 \AA. van der Waals and direct electrostatic
interactions were truncated at the cutoff. The direct sum non-bonded list was extended to cutoff + ``nonbond skin'' (10 + 2 \AA).\par
The solvated systems with harmonic restraints (force constant of 500 kcal/mol/\AA$^{2}$) on the position of each atom of
the protein were first subjected to 2000 steps of steepest descent energy minimization, followed by 1000 steps of conjugate gradient
minimization to remove bad contacts present in the initially built systems. The restraints on the protein atoms were sequentially
decreased to zero during further 4000 steps of energy minimization. The energy minimized systems were then slowly heated from 10 K
to the desired temperature in many steps during the first 210 ps of MD simulation. During this time, the solute particles were
restrained to their initial positions using harmonic restraints with a force constant of 20 kcal/mol/\AA$^{2}$. The
first 2 ns of equilibration simulations were performed in the NPT ensemble to attain the proper water density. Simulations
were then switched to the NVT ensemble for further production runs of 200--1450 ns,
depending on cation types.\par
%
\subsection*{Data Analysis}
All the analyses were carried out by using home-written codes and/or the AMBER17 tools.\cite{amber17tools} Images were rendered
using the Visual Molecular Dynamics software.\cite{humphrey1996vmd}\par
The free energy of ion binding, $\Delta G_{b}$, was calculated using the expression 
\begin{equation}\label{eq:FreeEne}
\Delta G_{b}=-k_{B}T \ln(C_{bI}/C_{fI}),
\end{equation}
where $k_B$ is the Boltzmann constant, and $C_{bI}$ and $C_{fI}$ are the concentration of bound and free ions, respectively. 
The expressions for calculations of the concentrations are $C_{bI}=N_{bI}/V_{s}$ and $C_{fI}=N_{fI}/V_{f}$, where $V_{s}$ is the
volume of the shell around the protein surface where ions are considered as bound, $N_{fI}$ (= total number of ions $-N_{bI}$)
is the number of free ions, and $V_{f}$ is the free volume available for ions.
The volumes were calculated following the protocol described in Ref. \cite{becconi2017protein}, by using the Gromacs
program \textit{gmx sasa}.\cite{abraham2015gromacs} Further details on the volume calculation are provided in the SI, section 3.
The last 200 ns data for each ion type (the last 150 ns for Na$^+$) was taken for the calculation of  $\Delta G_{b}$,
whereas the rest of the data served for the equilibration.\par
The reported entropy contributions in Figures \ref{fig:Entropy} and S6 were obtained by calculating absolute molar entropies for
free and protein-bound ions, free and protein-bound water molecules, and water molecules in the first and second SS$'s$ of the 
cation by using the 2PT method.\cite{lin2003two, lin2010two, pascal2011thermodynamics}
To generate trajectories for 2PT calculations, simulations were restarted after 100--500 ns, depending on the ion type.
3 short (40 ps) NVT trajectories for each system at each temperature were generated with coordinates and velocities saved every 4 fs.
To calculate the 2PT entropy for bound ions, we performed the analysis for all bound ions and got the average entropy per ion, similarly
for bound water. In Figures \ref{fig:Entropy} and S6, each point and the corresponding error bar are the average and standard deviation of
3 different simulations, respectively.\par
The surface or $\zeta$-potential was obtained by calculating as a function of $r$ (the distance from the center of mass of the protein)
the electrostatic potential profile, $\phi(r)$, for the system as follows.\cite{maiti2008counterion} $\phi(r)$ was calculated by solving
the Poisson equation, i.e, by carrying out a double integration of the charge density profile, $\rho (r)$, obtained from our MD
simulation by using the following expression.
\begin{equation}\label{eq:ZetaPot}
\phi(R)-\phi(r)=-\frac{1}{\epsilon}\int_{r}^{R} dr_1 \frac{1}{r_1^2} \int_{0}^{r_1} dr_2 r_2^2 \rho (r_2),
\end{equation}
where $R$ (= 65 \AA) is the radius of the inscribed sphere within the rectangular MD simulation box and $\epsilon$ is
the dielectric permittivity of water. A derivation of Eq. \ref{eq:ZetaPot} is given in the SI, section 4.
At a temperature $T$, $\epsilon$ was calculated from the Bjerrum length, $\lambda_B$, of water (= 7 \AA) by using the relation:
$\epsilon = e^2/4\pi\lambda_B k_BT$, where $e$ is the elementary charge. 
It should, however, be noted that for the temperature behavior of $\epsilon \propto T^{-\alpha}$, the exponent $\alpha$ in
experiments is close to $^3/_2$ but we take it to be 1 which is close to what is seen in simulations for a rather similar
water model.\cite{sedlmeier2013solvation} So the above approximation for $\epsilon(T)$ is deemed to be good for our purpose.
Finally, the $\zeta$-potential was obtained as $\zeta=\phi(R)-\phi(R_h+2r_c)$. Here, the hydrodynamic radius of the protein,
$R_h$, was taken to be 36 \AA{},\cite{li2012spectroscopic} and $r_c$$'s$ for the different cations (Figure \ref{fig:Sim1BSA}B) were taken
to be 2.8 \AA{} (Na$^{+}$), 2.7 \AA{} (Ca$^{2+}$), 2.3 \AA{} (Mg$^{2+}$) and 2.5 \AA{} (Y$^{3+}$).\par
$N_{Contacts}$ shown in Figure \ref{fig:Sim2BSA}B is defined as the total number of inter-protein amino acid residue--residue contacts,
and such a contact is counted if at least one pair of atoms from residues of two different proteins are within 3 \AA{}.\par 
$N_{Ion Bridges}$ shown in Figure \ref{fig:Sim2BSA}C is defined as the total number of ions bridging two different proteins, and an ion bridge
is counted if an ion is present within 3 \AA{} from both the proteins' surfaces.\par 
The average orientational autocorrelation function shown in Figure \ref{fig:Sim2BSA}D is defined as
\begin{equation}\label{eq:OrientCorr}
C(t)=\frac{1}{3} \sum_{i=1}^{3} \frac{\langle \cos\theta_{i}(0) \cdot \cos\theta_{i}(t) \rangle}{\langle \cos\theta_{i}(0) \cdot \cos\theta_{i}(0) \rangle},
\end{equation}
with $\cos\theta_{i}(t)= \hat{e}_{i}^{A} \cdot \hat{e}_{i}^{B} $, where $\hat{e}_{i}^{A}$ and $\hat{e}_{i}^{B}$ are the unit vectors
along the principal axes of proteins A and B, respectively, and the angular bracket represents the average over time origins.\par
%

\section*{Conflicts of interest}
There are no conflicts to declare.

\section*{Acknowledgements}
We thank Profs. Daan Frenkel and Tod Pascal for helpful discussions. We acknowledge
SERC, IISc for the allocation of computing time at the SAHASRAT machine. A.K.S. thanks MHRD, India for
the research fellowship and the Max Planck Society for financial support via the MaxWater initiative. 
R.R.N. thanks Infosys Foundation for support during his stay at IISc, Bangalore.



\balance


\end{document}


\clearpage

\begin{figure}
\centering
\includegraphics[width=0.45\textwidth]{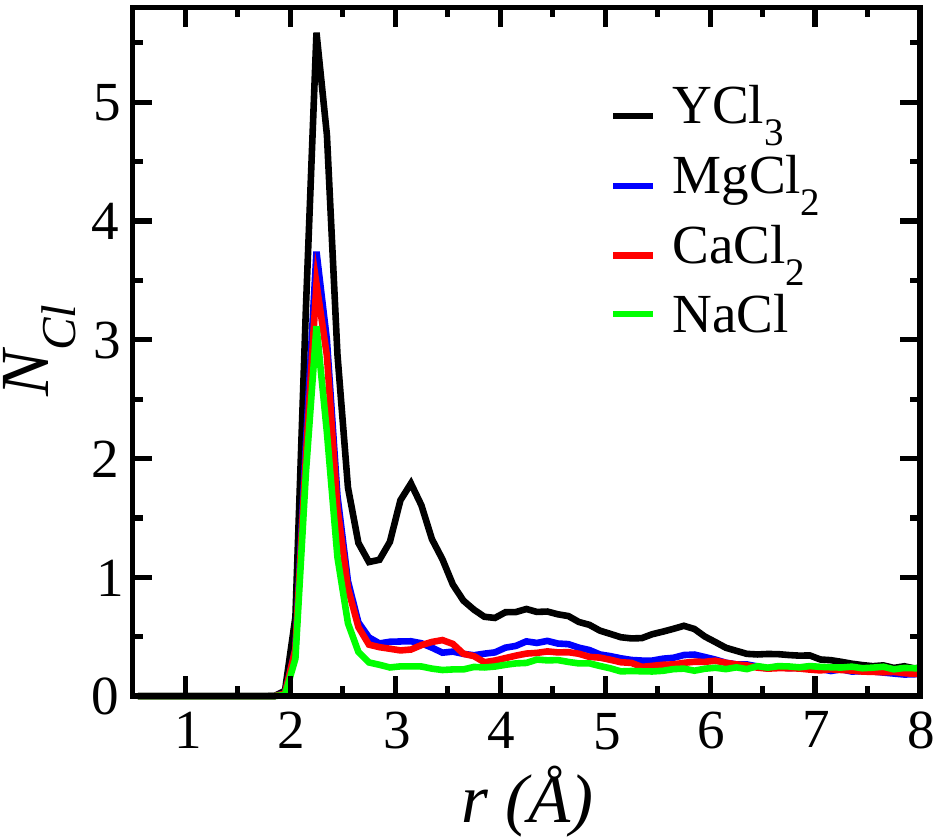}
\caption{For each of the different ionic solutions, the total number of Cl$^-$ ions ($N_{Cl}$) found within a shell of
width $dr$ $=0.1$ \AA, present at a shortest distance $r$ from the protein surface, is shown as a function of $r$
for the simulation performed at 303 K. $N_{Cl}(r)$ for each ionic solution type is averaged over the last 100 ns of the
simulation time.}
\end{figure}
\clearpage
\begin{figure}
\centering
\includegraphics[width=.65\textwidth]{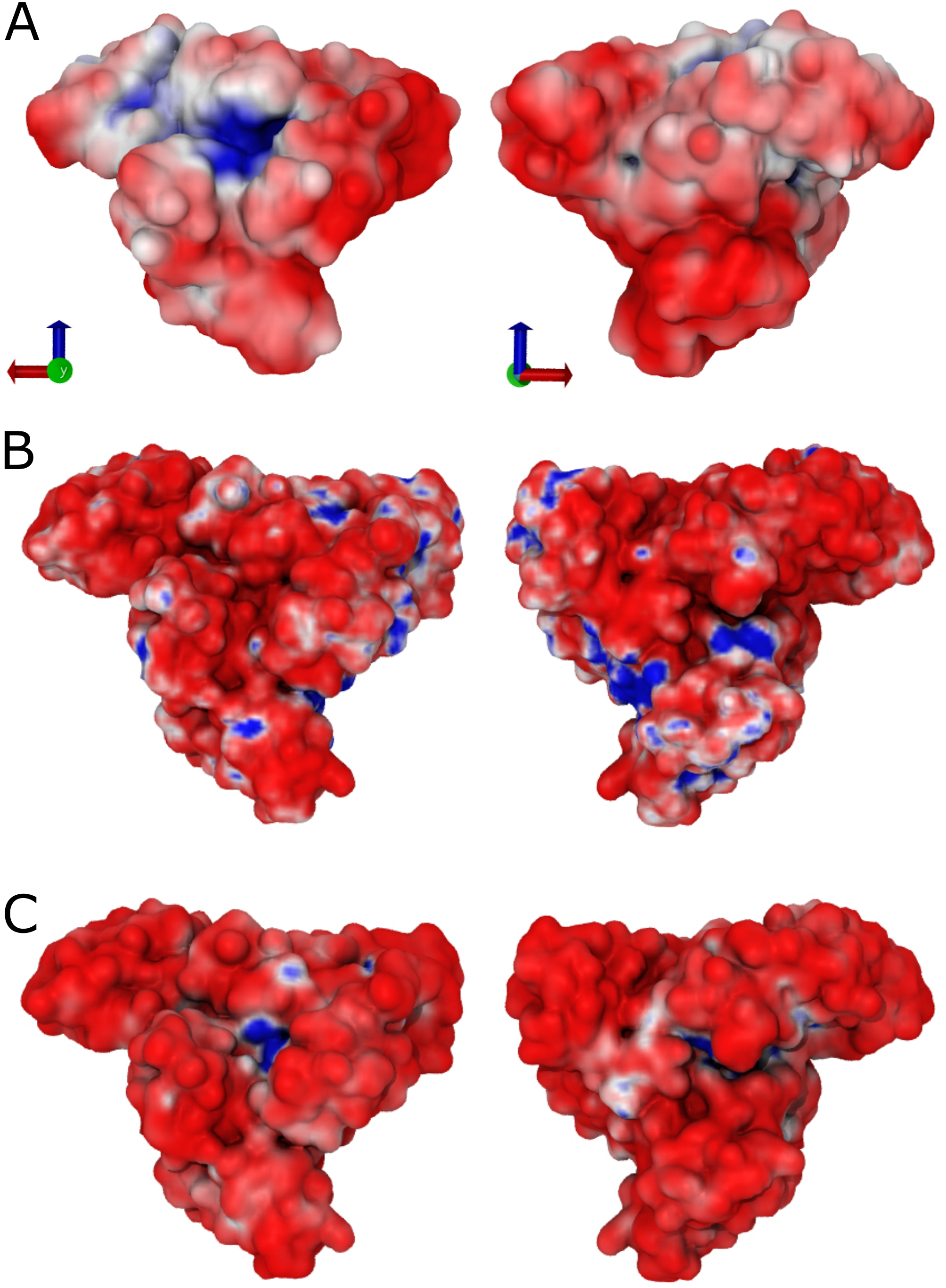}
\caption{(A) Electrostatic potential map for BSA, left: front view, right: back view. Red represents negative
potential, whereas blue represents positive potential. (B) Na$^+$ and (C) Cl$^-$ number density map obtained from
ion distributions within a 5 \AA{} shell from the protein surface sampled in the last 100 ns of the simulation at 303 K.
Red represents lower density, whereas blue represents higher density.}
\end{figure}
\clearpage
\begin{figure}
\centering
\includegraphics[width=0.45\textwidth]{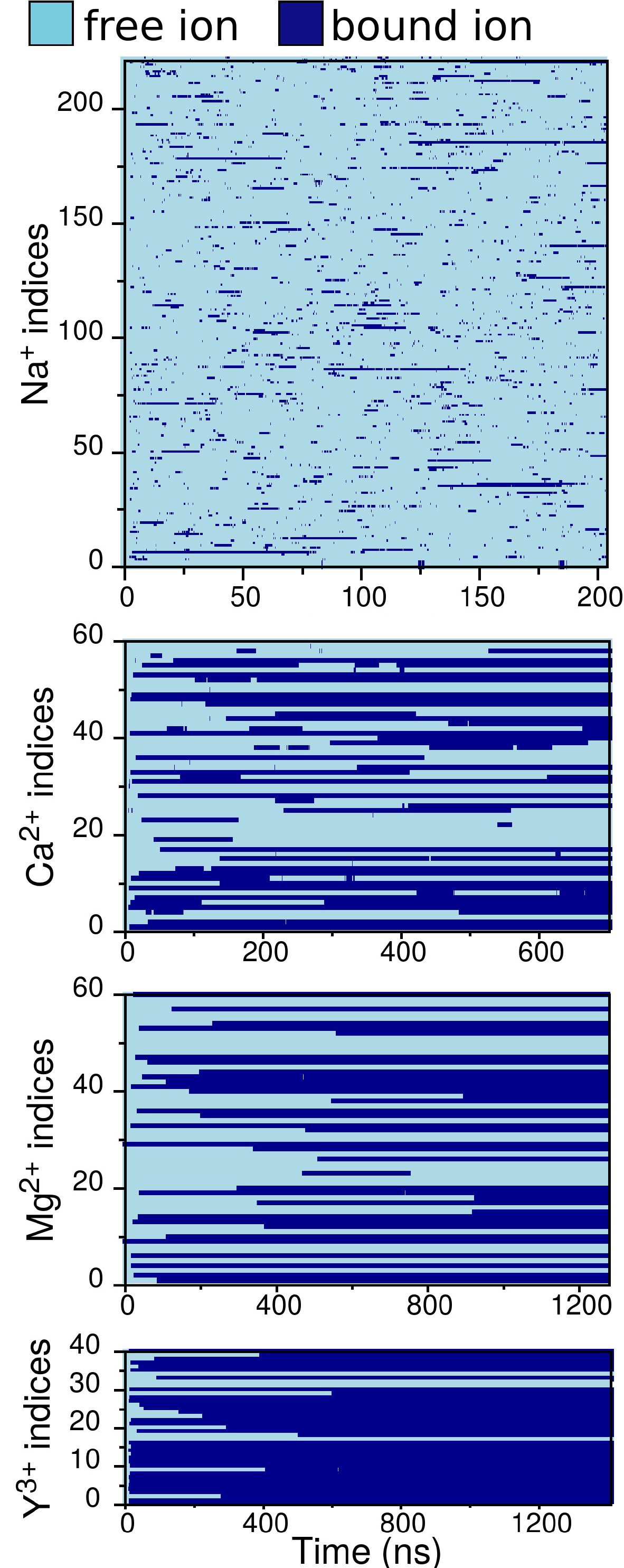}
\caption{Time series showing events of cation binding and unbinding from the protein surface at 303 K for each Na$^{+}$,
Ca$^{2+}$, Mg$^{2+}$, and Y$^{3+}$. The length of a continuous dark-blue (light-blue) line represents the
time for which a cation remains bound (unbound) to the protein surface.}
\end{figure}
\clearpage
\begin{figure}
\centering
\includegraphics[width=.88\textwidth]{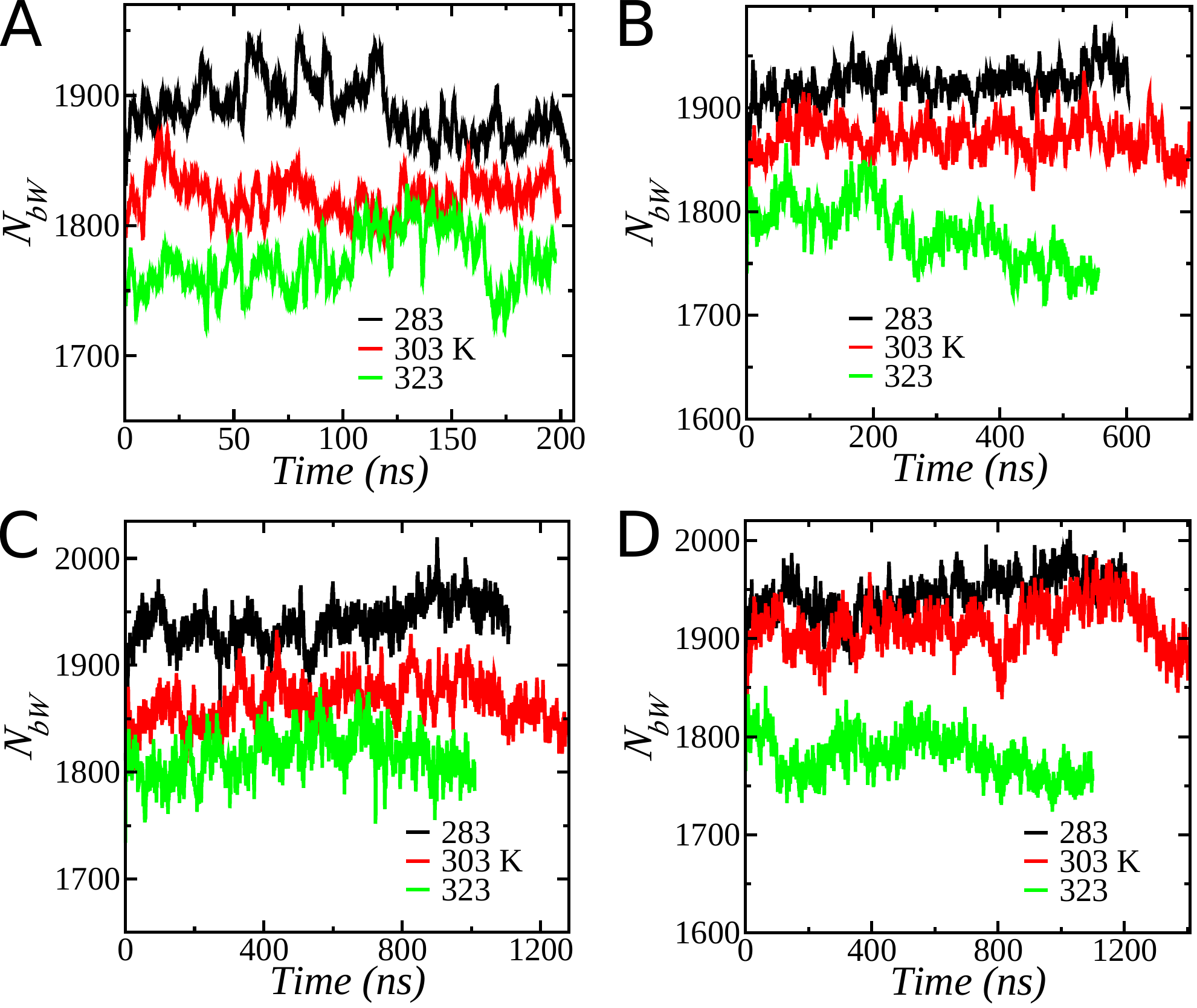}
\caption {Time series of the number of protein-bound water molecules at different temperatures for NaCl (A), CaCl$_2$ (B), MgCl$_2$
(C), and YCl$_3$ (D) solutions.}
\end{figure}
\clearpage
\begin{figure}
\centering
\includegraphics[width=.84\textwidth]{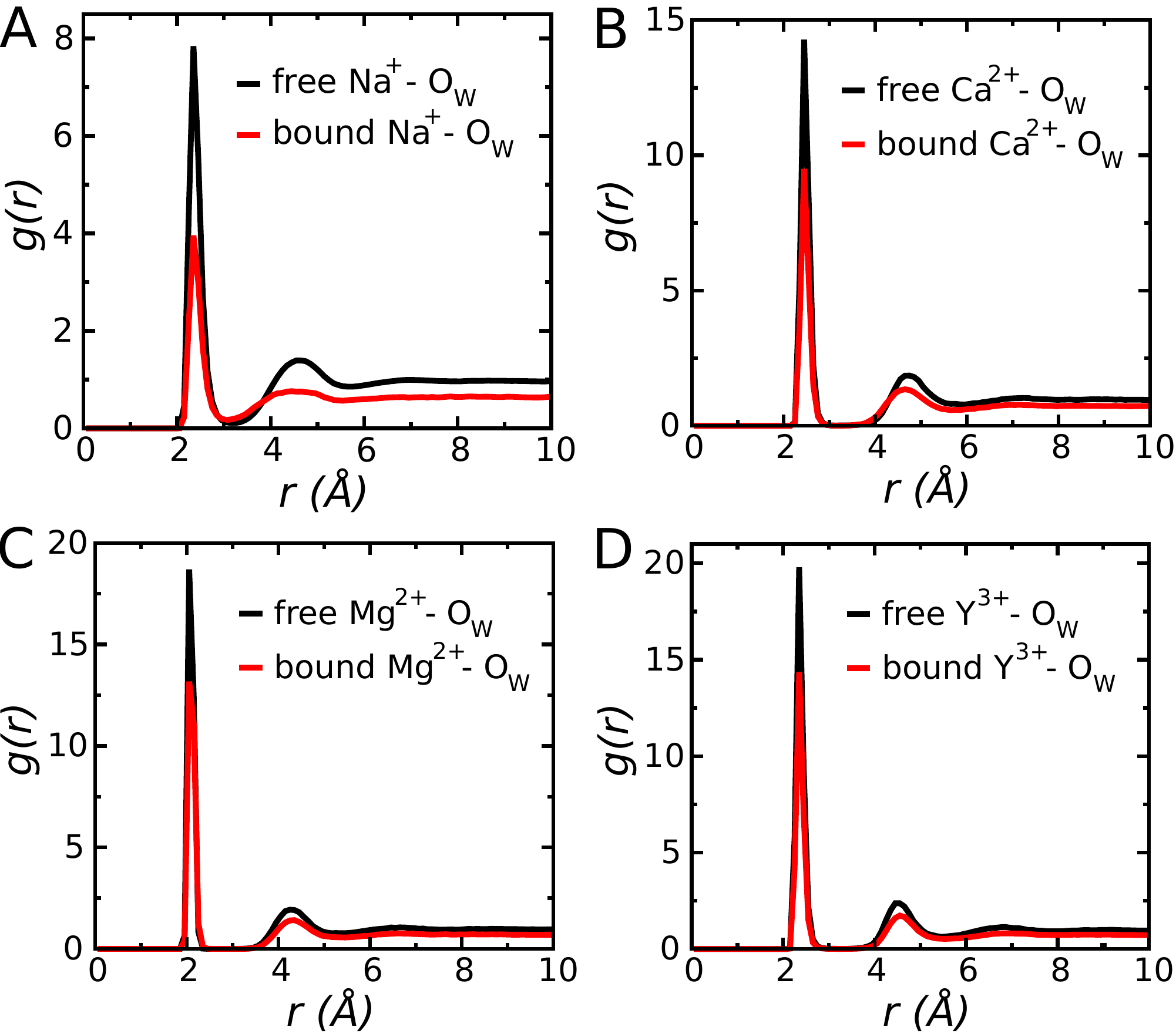}
\caption{The radial distribution functions (RDF) at 303 K for oxygen atoms of water around Na$^{+}$ (A), Ca$^{2+}$ (B), Mg$^{2+}$
(C), and Y$^{3+}$ (D) ions. In each case, the results for the cation free in solution and the cation bound to the protein surface
are shown. Water molecules are released from both the first and second solvation shells of each cation, when the cation 
binds to the protein surface.}
\end{figure}
\clearpage
\begin{figure}
\centering
\includegraphics[width=.45\textwidth]{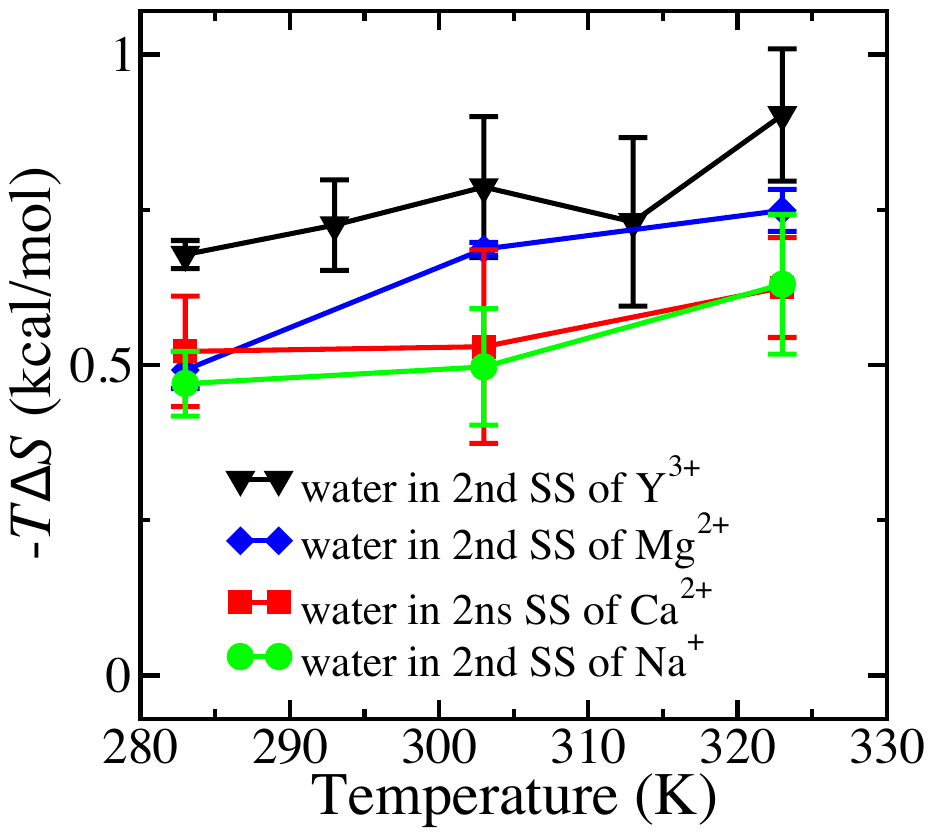}
\caption{Temperature dependence of the difference in entropies of a water in the second solvation shell of a cation and a water in bulk. Error bars represent the standard deviation. The lines are for guiding the eye.}
\end{figure}
\clearpage
\begin{figure}
\centering
\includegraphics[width=.45\textwidth]{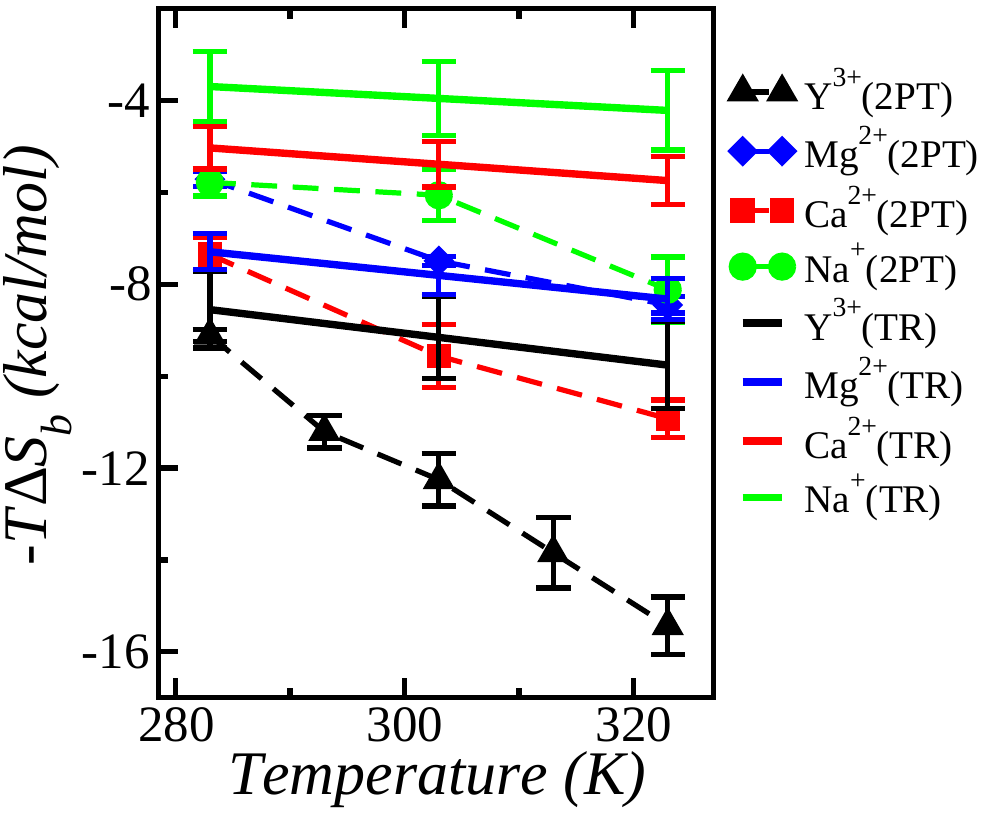}
\caption{Comparison of the total entropy contribution ($-T\Delta S_{b}$) obtained from the 2PT method and that obtained from
the temperature dependence of the ion-binding free energy $\Delta G_{b}$ (see Fig. 3A in the main text) using
the thermodynamic relation (TR), $\Delta S_{b}=-\partial \Delta G_{b}/\partial T$. The error bar for the 2PT method
represents the standard deviation of four independent calculations at each temperature, whereas the error bar for TR denotes
the standard linear-regression error associated with the estimation of the slope of temperature versus binding energy data.
Note that for each cation type, the entropy obtained from each method is negative and sufficiently large to drive the ion binding.
Given the higher ``true'' error (compared to the computed error here) usually associated with the entropy obtained from TR,
we deem the agreement between the widely different methods sufficient for our purposes.}
\end{figure}
\clearpage
\begin{figure}
\centering
\includegraphics[width=.45\textwidth]{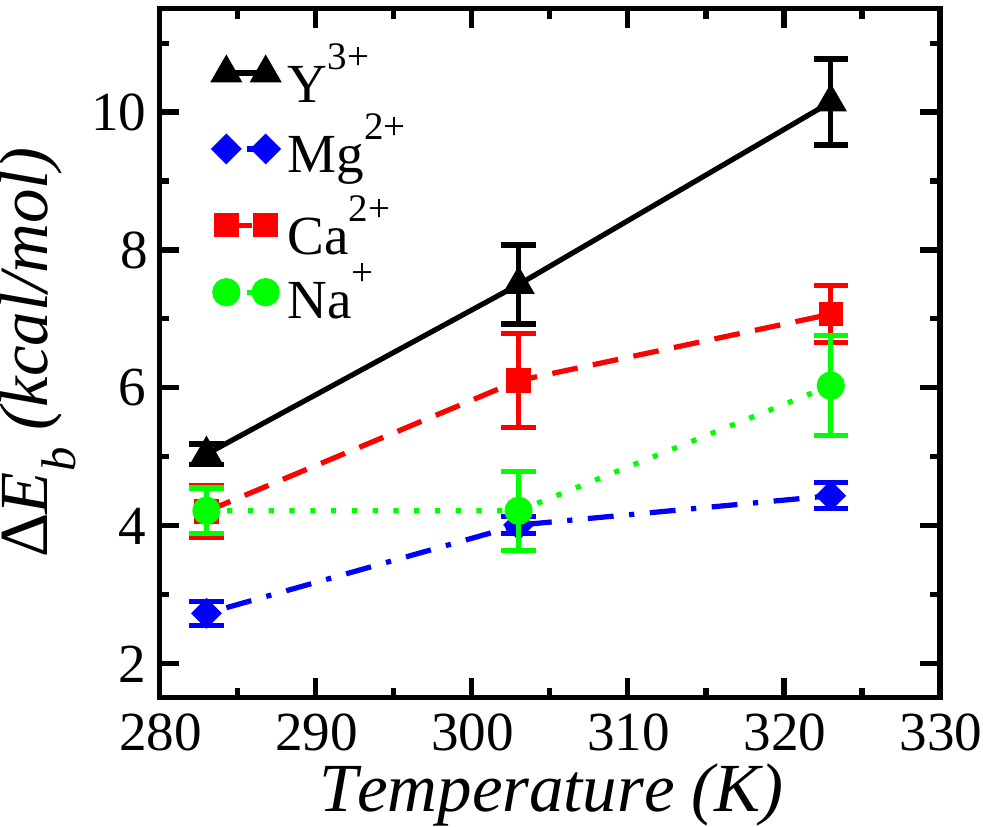}
\caption{Temperature dependence of the protein--ion binding energy $\Delta E_{b}$ for the different cations. The different lines are for guiding the eye. The error bars are the propagation errors coming from the errors in $\Delta G_{b}$ and $\Delta S_{b}$ (see Eq. 1 in the main text).}
\end{figure}
\clearpage
\begin{figure}
\centering
\includegraphics[width=.55\textwidth]{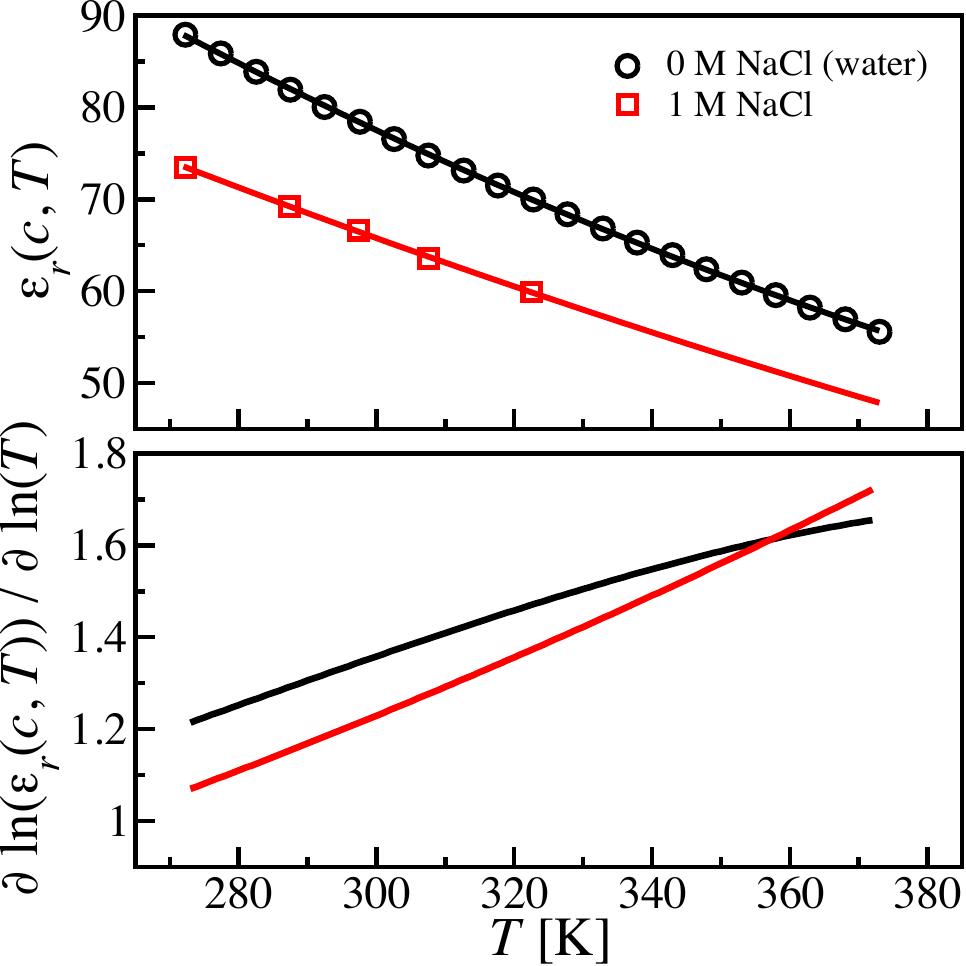}
\caption{(top) Temperature dependence of the relative dielectric constant $\epsilon_r$ of NaCl solutions.
Open circles are experimental data for pure water,\cite{robinson2002electrolyte} whereas open squares are experimental data for 1 M NaCl
solution.\cite{giese1970permittivity} Curves represent polynomial fits to the data. (bottom) First logarithmic derivative of $\epsilon_r$
which determines the exponent $\alpha$ of the scaling law: $\epsilon_r (c,T) \propto T^{-\alpha (c)}$.}
\end{figure}
\clearpage
\begin{figure}
\centering
\includegraphics[width=.90\textwidth]{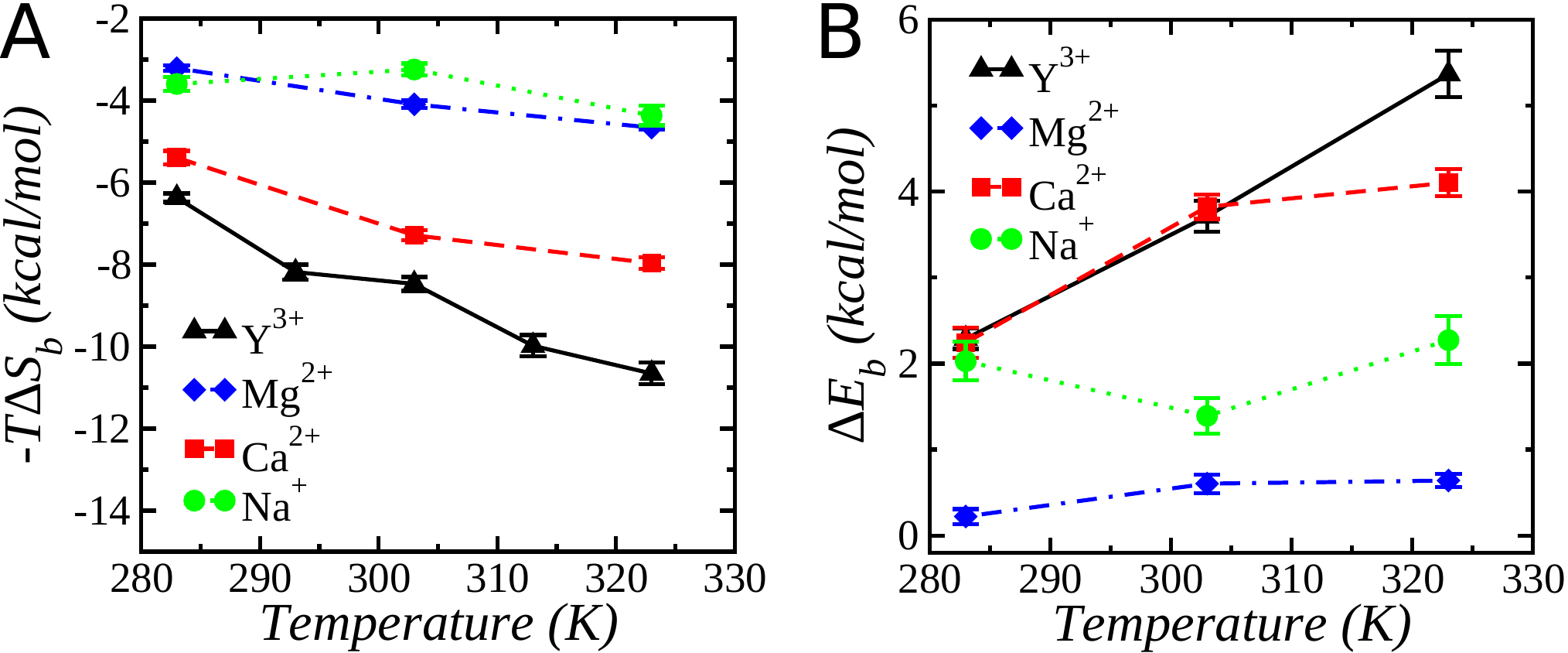}
\caption{Temperature dependence of the total entropy contribution $-T\Delta S_{b}$ (A) and the total energy contribution $\Delta E_{b}$ (B) for the different cations, where the respective contributions due to the dehydration of the second solvation shell are not included. The different lines are for guiding the eye. For comparison where all the contributions are included, see Fig. 3B (in the main text) and Fig. S8 for $-T\Delta S_{b}$ and $\Delta E_{b}$, respectively.}
\end{figure}
\clearpage

\section*{1. Temperature dependence of the free energy}
The free energy is given by 
\begin{equation*} 
	F=U-TS=U(T)-TS(T)=F(T,V), 
\end{equation*}
where $U$ is the internal energy, $S$ is the entropy, $V$ is the volume, and T is the temperature. 
Taking the derivative of $F$\\
\begin{equation*}
	\implies \frac{\partial F(T,V)}{\partial T}\bigg |_V=-S(T)=U'(T)-S(T)-TS'(T).
\end{equation*}
In the above equation, we have used the relation: $U'(T)=C_V=TS'(T)$, where $C_V$ is the specific heat at constant $V$.
Therefore, the change of $U$ with $T$ does not enter the $T$-dependence of $F$.\par 
%
\section*{2. Two-phase thermodynamic (2PT) method for entropy calculation}
The 2PT method was developed by Lin \textit{et al.}\cite{lin2003two, lin2010two} The central hypothesis of the 2PT method is that the density of
states (DoS) of a fluid can be treated as a combination of gas and solid-like components. The DoS of a fluid has a zero-frequency diffusive
mode $S(0)$, similar to a gas, and a maximum at some finite frequency followed by an exponential decay at higher frequencies, similar to a solid.\par
Lin \textit{et al.}\cite{lin2003two} showed that the thermodynamic properties can be estimated by treating the DoS
of a fluid as a sum of solid-like ($S^s(\nu)$) and gas-like ($S^g(\nu)$) contributions. Thermodynamic quantities for a solid can
be estimated by treating its vibrational modes as a system of noninteracting harmonic oscillators, as in the Debye
model \cite{mcquarrie1976statistical}. The gas part is described as a low-density hard-sphere fluid. The velocity autocorrelation
function decays exponentially for this model \cite{mcquarrie1976statistical}, and hence the DoS can be calculated analytically.
Thus, the calculation of entropy for solid and gas requires knowledge of the DoS.\par
The translational density of states $S(\nu)$ of a system is defined as the mass-weighted sum of atomic spectral densities
$s_j^k(\nu)$
\begin{equation}\label{eq:2pt1}
S(\nu) = \frac{2}{k_BT} \sum_{j=1}^{N}\sum_{k=1}^{3} {m_js_j^k(\nu)}\,,
\end{equation}
where $m_j$ is the mass of the $j$th atom, $k$ refers to the three Cartesian directions,
and $s_j^k(\nu)$ is given by:
\begin{equation}\label{eq:2pt2}
s_j^k(\nu) = \lim_{\tau\rightarrow\infty}\frac{\left|\int_{-\tau}^{\tau}\nu_j^k(t)e^{-i2\pi\nu t}dt\right|^2}{\int_{-\tau}^{\tau}dt} = \lim_{\tau\rightarrow\infty}\frac{\left|\int_{-\tau}^{\tau}\nu_j^k(t)e^{-i2\pi\nu t}dt\right|^2}{2\tau}\,.
\end{equation}
Here, $\nu_j^k(t)$ is the $k$th component of the velocity of atom $j$. It can be shown that
the atomic spectral density $s_j^k(\nu)$ can be obtained from the Fourier transform
of the velocity auto-correlation function (VACF) $c_j^k(t)$\cite{lin2003two}
\begin{equation}\label{eq:2pt3}
s_j^k(\nu) = \lim_{\tau\rightarrow\infty} \int_{-\tau}^{\tau}{c_j^k(t)e^{-i2\pi\nu t}dt}\,,
\end{equation}
where $c_j^k(t)$ is given by:
\begin{equation}\label{eq:2pt4}
c_j^k(t) = \lim_{\tau\rightarrow\infty} \frac{1}{2\tau} \int_{-\tau}^{\tau}{\nu_j^k(t+t^\prime)\nu_j^k(t^\prime)dt^\prime}\,.
\end{equation}
Thus, Eq. \ref{eq:2pt1} can be rewritten as:
\begin{equation}\label{eq:2pt5}
S(\nu) = \frac{2}{k_BT} \lim_{\tau\rightarrow\infty} \int_{-\tau}^{\tau} {\sum_{j=1}^{N}\sum_{k=1}^{3} {m_jc_j^k(t)e^{-i2 \pi \nu t}dt} }\,.
\end{equation}
More generally, it can be written as:
\begin{equation}\label{eq:2pt6}
S(\nu) = \frac{2}{k_BT} \lim_{\tau\rightarrow\infty} \int_{-\tau}^{\tau} {C(t)e^{-i2 \pi \nu t}dt}\,.
\end{equation}
In the above equation, $C(t)$ can be either the mass-weighted translational VACF determined
from the center of mass velocity $V_i^{cm}(t)$ of the $i$th molecule,
\begin{equation}\label{eq:2pt7}
C_T(t) = \sum_{i=1}^{N} {\left<m_i V_i^{cm}(t).V_i^{cm}(0)\right>}
\end{equation}
or the moment-of-inertia weighted angular velocity auto-correlation function
\begin{equation}\label{eq:2pt8}
C_R(t) = \sum_{i=1}^{3} \sum_{i=1}^{N} {\left<I_{ij}\omega_{ij}(t)\omega_{ij}(0)\right>}\,,
\end{equation}
where $I_{ij}$ and $\omega_{ij}$ are the $j$th components of the moment of inertia tensor and the angular velocity of
the $i$th molecule, respectively. One can obtain the translational or rotational DoS depending on the use of $C_T(t)$
or $C_R(t)$ in Eq. \ref{eq:2pt6}.\par
In the 2PT method, the DoS is decomposed into a gas-like diffusive component and a solid-like nondiffusive component,
$S(\nu) = S^g(\nu) + S^s(\nu)$, using the fluidity factor $f$ which is a measure of the fluidity of a system. $f$
is estimated in terms of the dimensionless diffusivity $\Delta$ using the universal equation \cite{lin2003two}:
\begin{equation}\label{eq:2pt9}
2\Delta^{-9/2}f^{15/2}-6\Delta^{-3}f^{5}-\Delta^{-3/2}f^{7/2}+6\Delta^{-3/2}f^{5/2}+2f-2=0\,.
\end{equation}
The diffusivity $\Delta$ can be uniquely determined for a thermodynamic state of the system using the equation:
\begin{equation}\label{eq:2pt10}
\Delta(T,\rho,m,S_0) = \frac{2S_0}{9N}\left(\frac{6}{\pi}\right)^{2/3}\left(\frac{\pi k_BT}{m}\right)^{1/2}\rho^{1/3}\,,
\end{equation}
where $S_0 = S(0)$ is the zero-frequency component of the DoS function (translational or rotational). Knowing $f$ from
Eqs. \ref{eq:2pt9} and \ref{eq:2pt10}, the gas-like diffusive component of the DoS can be obtained using a
hard-sphere diffusive model:
\begin{equation}\label{eq:2pt11}
S^g(\nu) = \frac{S_0}{1+\left[\frac{\pi S_0\nu}{6fN}\right]^2}\,.
\end{equation}
Lin \textit{et al.}\cite{lin2003two} used the gas--solid decomposition scheme only for the translational DoS of monoatomic fluids.
In a later work, Lin \textit{et al.}\cite{lin2010two} showed that for polyatomic fluids, the rotational entropy can be computed
if the decomposition scheme is used for the rotational DoS as well. Separate fluidity factors $f$s are determined for the
translational and rotational DoS using the translational and rotational diffusivities in Eq. \ref{eq:2pt9}. Then, the gas-like
component of entropy is calculated using Eq. \ref{eq:2pt11} with $S_0$ being $S_{tran}(0)$ or $S_{rot}(0)$ for the translational
and rotational cases, respectively. Once such decomposition of DoS is done, each thermodynamic quantity $A_m$ can be computed
from the solid-like and gas-like DoS functions with the corresponding weight functions as follows.
\begin{equation}\label{eq:2pt12}
A_m = \beta^{-1}\left[\int_{0}^{\infty}{d\nu S_m^g(\nu)W_{A,m}^{g}} + \int_{0}^{\infty}{d\nu S_m^s(\nu)W_{A,m}^{s}}\right]\,,
\end{equation}
where $m$ can be translational, rotational or vibrational. The weight functions are provided in Ref. \cite{lin2010two}.
For the rigid TIP3P \cite{jorgensen1983comparison} water model used in our simulations, the contribution due to intra-molecular
vibration is zero.
%
\section*{3. Definition of volume for the calculation of ion concentration}
Following Ref. \cite{becconi2017protein}, the volume of the shell surrounding the protein surface ($V_s$) where ions are considered as bound
is defined as $V_s=V_{ps}-V_{prot}$. $V_{prot}$ is the volume of the protein calculated by rolling a small probe sphere of radius 0.5 \AA{}
on the protein.
$V_{ps}$ is the volume occupied by the protein and the shell around the protein that contains the bound ions, and $V_{ps}$ is calculated 
by rolling around the protein surface a probe sphere of radius $r_{ps}$, which is equivalent to the contact distance of a bound ion from 
the protein surface.
The volume available for free ions ($V_f$) is defined as $V_f=V_{box}-V_{ps}$, where $V_{box}$ is the volume of the simulation box. 
Note that the calculation of $V_{ps}$ using this method is \textit{ad hoc} and influences the concentration of bound and free ions,
and hence the value of $\Delta G_b$ (\textit{cf.} Eq. 3 in the main text). We, therefore, varied $r_{ps}$ to get an optimized probe radius,
such that the calculated $\Delta G_b$ matches the experimental $\Delta G_b$ for Y$^{3+}$, as shown in the figure below.\par
\begin{figure}
\centering
\includegraphics[width=0.45\textwidth]{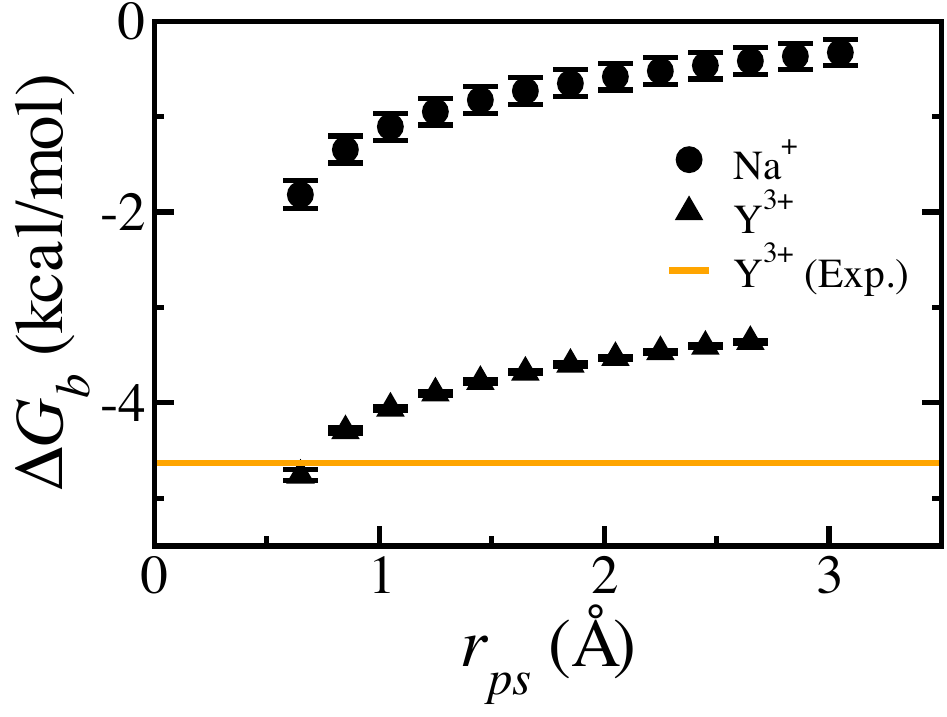}
\caption{Dependence of the binding free energy $\Delta G_b$ on the probe radius $r_{ps}$ used for defining the shell volume $V_s$.
The experimental data for Y$^{3+}$ is taken from Matsarskaia \textit{et al.}\cite{matsarskaia2016cation}}
\end{figure}
%
\section*{4. The surface or $\zeta$-potential calculation from the simulation data}
For the estimation of the surface or $\zeta$-potential of the protein in a salt solution, we consider Poisson's equation
\begin{equation}\label{eq:ZetaPot1}
	\nabla^{2}\phi=-\frac{\rho}{\epsilon},
\end{equation}
in \textit{spherical polar} coordinates. Here, $\phi$ is the electrostatic potential, $\rho$ is the charge density, and $\epsilon$ is
the dielectric permittivity of water. For simplicity, we assume that the protein is a spherically isotropic object. To justify this,
we calculate the asphericity, $\delta$, of BSA protein ($\delta=1-3[I_{x}I_{y}+I_{y}I_{z}+I_{x}I_{z}]/[I_{x}+I_{y}+I_{z}]^2$,
where $I_x$, $I_y$, and $I_z$ are the principal moments of inertia) and find $\delta$ to be very small ($=0.0196\pm0.0016$). 
Now, $\phi$ and $\rho$ are functions of only the radial distance $r$. Under these considerations, Eq. \ref{eq:ZetaPot1} becomes
\begin{equation}\label{eq:ZetaPot2}
	\frac{1}{r^2}\frac{\partial }{\partial r}\left(r^2\frac{\partial }{\partial r}\right)\phi(r)=-\frac{\rho(r)}{\epsilon}
\end{equation}
\begin{equation}\label{eq:ZetaPot3}
	\implies \frac{\partial }{\partial r}\left(r^2E(r)\right)=\frac{\rho(r)}{\epsilon}r^2,
\end{equation}
where the electric field $E=-\frac{\partial \phi(r)}{\partial r}$. Integrating both sides of the above equation from 0 to $r_1$
\begin{equation}\label{eq:ZetaPot4}
	r_{1}^2E(r_1)=\frac{1}{\epsilon}\int\limits_{0}^{r_1}dr_2\rho(r_2)r_{2}^2
\end{equation}
\begin{equation}\label{eq:ZetaPot5}
	\implies -\frac{\partial \phi(r)}{\partial r}=\frac{1}{\epsilon}\frac{1}{r_{1}^2}\int\limits_{0}^{r_1}dr_2\rho(r_2)r_{2}^2.
\end{equation}
To obtain the electrostatic potential profile $\phi(r)$, integrating both sides of the above equation from r to $R$ (the maximum
radial distance possible due to the finite size of the simulation box under \textit{periodic boundary condition})
\begin{equation}\label{eq:ZetaPot6}
	\implies \phi(R)-\phi(r)=-\frac{1}{\epsilon}\int\limits_{r}^{R}dr_1\frac{1}{r_{1}^2}\int\limits_{0}^{r_1}dr_2\rho(r_2)r_{2}^2.
\end{equation}
Note that this equation is the same as Eq. 9 in the main text. The surface or $\zeta$-potential of the protein is defined as
the electrostatic potential at one ionic diameter away from the protein surface. So, the $\zeta$-potential is evaluated from the above
equation using the charge density $\rho(r)$ obtained from the simulation as, $\zeta=\phi(R)-\phi(R_h+2r_c)$. Here, $R_h$ is the
hydrodynamic radius of the protein 
obtained to be 36 \AA{} from dynamic light scattering experiments,\cite{li2012spectroscopic} and $r_c$ is the
effective radius of the counterion. The values of the parameters used to obtain the $\zeta$-potential
are provided in the Methods section in the main text.
%

\begin{table}\centering
\caption{Structural parameters (such as the coordination number of the ion, $N_{hyd}$, and the ion–oxygen distance for
water present in the ion's 1$^{st}$ solvation shell, $d_{I-O}^{1st SS}$, and 2$^{nd}$ solvation shell $d_{I-O}^{2nd SS}$) and
entropy ($\Delta S_{hyd}$) of ion hydration obtained from simulations. The corresponding experimental values are given within the
brackets. The calculated values for $d_{I-O}^{1st SS}$, $d_{I-O}^{2nd SS}$, and $N_{hyd}$ are in quantitative
agreement with the corresponding experimental values \cite{marcus2009effect, jalilehvand2001hydration}. The computed values of
$\Delta S_{hyd}$ match well with the experimental values \cite{marcus2012ions} for all the ions, except for Na$^+$.
The $\sim$50\% overestimation in the calculated $\Delta S_{hyd}$ for Na$^{+}$ may be due to the inaccurate estimation of
the entropy of a water molecule present in the 2$^{nd}$ SS of Na$^+$. The lifetime of a water molecule in the 2$^{nd}$ SS of Na$^+$ ion is
10--15 ps. Within this short time period, the velocity--velocity autocorrelation---which is needed to obtain the spectral density-of-states
that serves as an input for the 2PT entropy calculation \cite{lin2003two, lin2010two}--- is not well converged.
}

\begin{tabular}{llllr}
	Ion & $d_{I-O}^{1st SS}$ (\AA) & $d_{I-O}^{2nd SS}$ (\AA) & $N_{hyd}$ & $\Delta S_{hyd}$ (J mol$^{-1}$ K$^{-1}$) \\
\hline
	Na$^{+}$ & 2.35$\pm$0.05 (2.34) & 4.55$\pm$0.05 (---) & 5.7 (5.6$\pm$0.3) & -168.08$\pm$20.94 (-111.2) \\
\hline
	Ca$^{2+}$ & 2.45$\pm$0.05 (2.46) & 4.65$\pm$0.05 (4.58) & 8.0 (8) & -270.57$\pm$39.89 (-252.4) \\
\hline
	Mg$^{2+}$ & 2.05$\pm$0.05 (2.09) & 4.25$\pm$0.05 (4.35) & 6.0 (6) & -316.09$\pm$2.72 (-331.2) \\
\hline
	Y$^{3+}$ & 2.35$\pm$0.05 (2.37) & 4.50$\pm$0.05 (4.40) & 9.0 (8) & -460.77$\pm$29.96 (-482.5) \\
\hline
	Cl$^{-}$ & 3.15$\pm$0.05 (3.14) & 5.05$\pm$0.05 (4.99) & 7.2 (7) & -71.56$\pm$9.43 (-75.7) \\
\end{tabular}
\end{table}

\begin{table}\centering
\caption{In the process of a cation binding to the protein, the average numbers of water molecules released from the protein surface and the first and second solvation shells (SS) of the cation are given at each temperature for the different cations.}

\begin{tabular}{llrrr}
	System & Temperature & Protein & 1$^{st}$ SS of cation & 2$^{nd}$ SS of cation \\
\hline
                & 283 & 4.37 & 2.47 & 4.65 \\
Protein in NaCl & 303 & 4.55 & 2.52 & 5.68 \\
                & 323 & 4.39 & 2.82 & 5.96 \\
\hline
                    & 283 & 4.74 & 2.32 & 3.76 \\
Protein in CaCl$_2$ & 303 & 5.63 & 2.62 & 4.30 \\
                    & 323 & 5.34 & 3.17 & 4.74 \\
\hline
                    & 283 & 2.68 & 1.04 & 5.08 \\
Protein in MgCl$_2$ & 303 & 2.90 & 1.15 & 4.95 \\
                    & 323 & 2.79 & 1.26 & 5.06 \\
\hline
                   & 283 & 4.16 & 2.09 & 4.05 \\
                   & 293 & 5.07 & 2.49 & 4.17 \\
Protein in YCl$_3$ & 303 & 5.10 & 2.47 & 4.81 \\
                   & 313 & 5.61 & 2.69 & 5.29 \\
                   & 323 & 5.40 & 2.87 & 5.29 \\
\end{tabular}
\end{table}
\clearpage
